\documentclass[
reprint,
superscriptaddress,
amsmath,
amssymb,
aps,
pra
]{revtex4-2}
\usepackage{tabularx}
\usepackage{adjustbox}
\usepackage{longtable}
\usepackage{makecell}  
\usepackage{graphicx}
\usepackage{dcolumn}
\usepackage{bm}
\usepackage[colorlinks, linkcolor=blue, anchorcolor=blue, citecolor=blue]{hyperref}
\usepackage{color}

\begin{document}

\preprint{APS/123-QED}

\title{High-dimentional Multipartite Entanglement Structure Detection with Low Cost}


\author{Rui Li}
\affiliation{School of Physics, Beihang University, Beijing 100191, China}

\author{Shikun Zhang}
\affiliation{School of Physics, Beihang University, Beijing 100191, China}

\author{Zheng Qin}
\affiliation{School of Physics, Beihang University, Beijing 100191, China}

\author{Chunxiao Du}
\affiliation{School of Physics, Beihang University, Beijing 100191, China}

\author{Yang Zhou}
\email[Corresponding author: ]{yangzhou9103@buaa.edu.cn}
\affiliation{School of Physics, Beihang University, Beijing 100191, China}

\author{Zhisong Xiao}
\affiliation{School of Physics, Beihang University, Beijing 100191, China}
\affiliation{School of Instrument Science and Opto-Electronics Engineering, Beijing Information Science and Technology University, Beijing 100192, China}

\date{\today}

\begin{abstract}
Quantum entanglement detection and characterization are crucial for various quantum information processes. Most existing methods for entanglement detection rely heavily on a complete description of the quantum state, which requires numerous measurements and complex setups. This makes these theoretically sound approaches costly and impractical, as the system size increases. 
In this work, we propose a multi-view neural network model to generate representations suitable for entanglement structure detection. The number of required quantum measurements is polynomial rather than exponential increase with the qubit number. This remarkable reduction in resource costs makes it possible to detect specific entanglement structures in large-scale systems. Numerical simulations show that our method achieves over 95\% detection accuracy for up to 19 qubits systems. By enabling a universal, flexible and resource-efficient analysis of entanglement structures, our approach enhances the capability of utilizing quantum states across a wide range of applications.
\end{abstract}

\maketitle


\section{\label{sec:level1}Introduction}

Quantum entanglement~\cite{einstein1935can,schrodinger1935discussion,chitambar2019quantum} describes the intricate correlations between subsystems that defy classical physics. This non-classical feature is central to quantum theory and has been intensely researched since its inception. Detecting and characterizing entanglement structures is crucial for both fundamental physics~\cite{PhysRevB.92.235116,PhysRevB.97.245135,laflorencie2016quantum} and various potential applications, including quantum computation~\cite{feynman1982simulating,zhang2024single,qin2024applicability}, quantum communication~\cite{alber2001mixed,ursin2007entanglement,pan2001entanglement}, and quantum metrology~\cite{pezze2018quantum,ren2021metrological}. An improper amount of entanglement, incorrect structure, or pattern of multipartite entangled states can affect the overall performance of the given quantum protocols~\cite{PhysRevLett.102.190501, Dür_2007}. As a result,  knowledge about the entangled states is necessary. Most existing methods for entanglement detection, whether it is analytic or data-driven, rely heavily on a complete description of the quantum state~\cite{RevModPhys.81.865,guhne2009entanglement,PhysRevLett.77.1413,zhou2019detecting,li2024entanglement, chen2021detecting,zhang2023entanglement}.

\begin{figure}[h]
\centering
\includegraphics[width=0.3\textwidth]{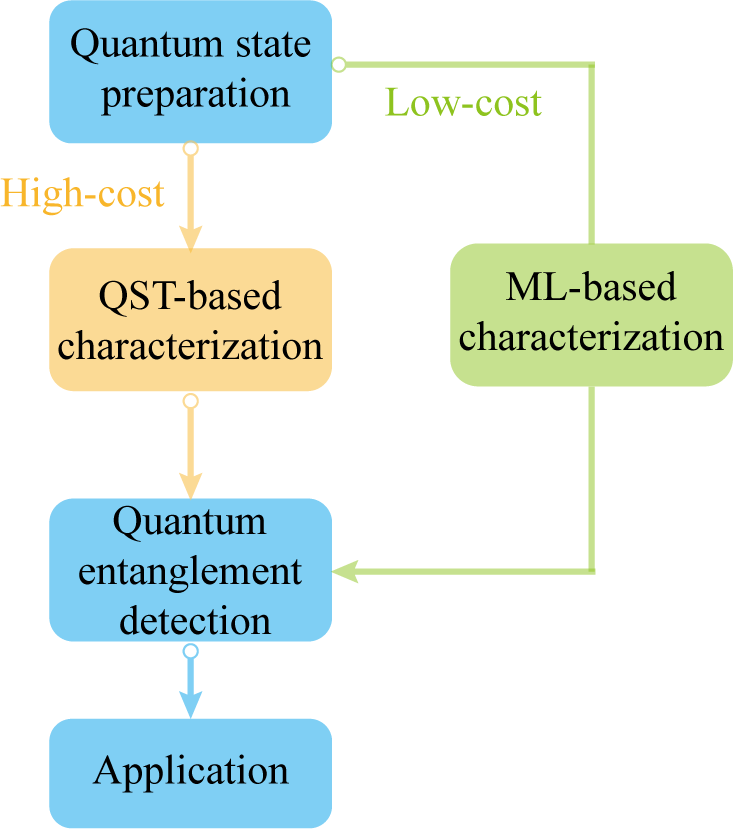}
\caption{Comparison of traditional high-cost QST-based approach vs. our low-cost ML-based method}
\label{figintro}
\end{figure}

As illustrated in Fig.~\ref{figintro}, a typical quantum information process (QIP) involves quantum state preparation, full quantum state tomography, and entanglement detection before application. Quantum state tomography (QST)~\cite{paris2004quantum,ahmed2021quantum,xin2017quantum} has been widely used as an experimental procedure for obtaining a complete quantum mechanical description of a system. It can reconstruct the density matrix of quantum state through measurements on an ensemble of identical quantum states. However, QST faces severe scalability issues as the number of qubits grows. For an $n$ qubits system, the conventional QST requires $4^n - 1$ independent measurements, which is prohibitively expensive for systems with more than a few qubits~\cite{paris2004quantum,gross2010quantum,Altepeter2004}. This exponential scaling is a fundamental bottleneck in characterizing quantum systems and significantly increases experimental and computational costs.

The challenges of QST have prompted researchers to explore more cost-effective approaches~\cite{torlai2018neural,gao2018experimental}. Recent advancements have shown that machine learning (ML) techniques can offer a natural and general way of performing QST from a limited amount of experimental data, circumventing the need for costly full state reconstruction~\cite{carleo2017solving,carrasquilla2020machine,dunjko2018machine}. However, representations generated in such a way are incompatible with most of the existing entanglement detection methods which take the density matrix as their starting point. This makes detecting entanglement structures in large-scale quantum systems remain challenging. Therefore, there is an urgent need to develop new techniques that can directly detect entanglement structures from only finite measurements.

In this work, we propose a low-cost approach based on ML to detect entanglement structures directly from limited measurement outcomes, bypassing the need for costly full quantum state tomography. Our method uses a multi-view representation learning framework~\cite{li2018survey,wang2015deep} to integrate information from various partial measurements of the quantum system and a neural network to classify various entanglement structures. By removing the necessity for full state tomography, our method achieves high-accuracy entanglement detection with significantly fewer resources than traditional QST-based approaches. For systems up to 19 qubits, we demonstrate a reduction of approximately three orders of magnitude in required measurements while maintaining detection fidelity above 95\%. This demonstrates that our approach can significantly reduce the number of required measurements, lower experimental costs, and extend the applicability of entanglement detection to larger quantum systems previously too costly to analyze.




It is noteworthy that entanglement witness (EW) ~\cite{guhne2009entanglement,lewenstein2000optimization} can also provide an economic method of detection that doesn't need the full information about the quantum state. Here, only the information about the mean value of some observable in a given quantum state is needed~\cite{RevModPhys.81.865}. However, EW is state-specific and can only be used to test whether it is separable or entangled. Our approach is different in that it can be applied to classify different entanglement structures with different qubit numbers. In the following sections, we detail our methodology, present comprehensive numerical results demonstrating its cost-effectiveness across various system sizes, and discuss the broader implications of our findings in quantum information science.

The structure of this paper is organized as follows. In Section \hyperref[sec:levelP]{II}, we introduce the concept of entanglement separability and entanglement depth, which are used as descriptors of the entanglement structure. In Section \hyperref[sec:levelM]{III}, we describe the dataset preparation process for our analysis and introduce a multi-view model that captures the inherent properties of entangled quantum states to detect and classify various entanglement structures. In Section \hyperref[sec:levelR]{IV}, we quantify the performance of the trained neural network with numerical experiments. Finally, this paper is concluded in Section \hyperref[sec:level4]{V}.

\section{\label{sec:levelP}PRELIMINARY}

\subsection{\label{sec:level2}Separable vs. entangled state}

For two quantum systems $A$ and $B$, their states are represented by the state vectors $\left.|{\mathrm{\Psi}}_A\right\rangle$ and $\left.|{\mathrm{\Psi}}_B\right\rangle$ in their Hilbert spaces $H_A$ and $H_B$. The joint quantum state of the two systems is represented by the state vector $\left.|{\mathrm{\Psi}}_{AB}\right\rangle= \left.|{\mathrm{\Psi}}_A\right\rangle\otimes\left.|{\mathrm{\Psi}}_B\right\rangle$ in the Hilbert space $H_A\otimes H_B$. If $\left.|{\mathrm{\Psi}}_{AB}\right\rangle$ can be written as a product state, then it is a separable state; otherwise, it is an entangled state. 

Generalized to an $N$-qubit case, the quantum state is usually represented by a density matrix:
\begin{equation}
    \hat{\rho} = \sum_i p_i\hat{\rho}_i.
    \label{eq2}
\end{equation}
Here, $p_i$ is the probability that $0\leq p_i\leq 1$ and $\sum_i p_i=1$, $\hat{\rho}_i^k $ is the pure state density matrix of each subsystem. If this density matrix $\rho$ can be expressed as a convex combination of multiple product states as follows,  
\begin{equation}
    \hat{\rho} = \sum_l p_i\hat{\rho}_i^1\otimes\hat{\rho}_i^2\ldots\otimes\hat{\rho}_i^N
    \label{eq3}
\end{equation}
the corresponding state is considered separable; otherwise, it is entangled.

To demonstrate this approach, we mainly consider the GHZ state, a paradigmatic N-qubit multipartite entangled quantum state defined as:
\begin{equation}
    \left.|{\mathrm{\Psi}}_{GHZ}\right\rangle= \frac{1}{\sqrt2}\ (\left.|1\right\rangle^ {\otimes N}+\left.|0\right\rangle^ {\otimes N})
    \label{eq4ghz}
\end{equation}

 In this work, GHZ states were generated using quantum circuits. It is started by initializing all qubits in the \(|0\rangle\) state. Then, a Hadamard gate (H gate) is applied to the first qubit, which transforms its state from \(|0\rangle\) to \(\frac{1}{\sqrt{2}}(|0\rangle + |1\rangle)\). Following this, Controlled-NOT (CNOT) gates are applied between the first qubit and each of the subsequent qubits. This sequence of operations entangles qubits, thereby creating the required GHZ state. An example with three qubits is illustrated in Fig.~\ref{fig2}(a).

\subsection{\label{sec3}Entanglement structure}

\begin{figure}[h]
    \centering
    \includegraphics[width=0.5\textwidth]{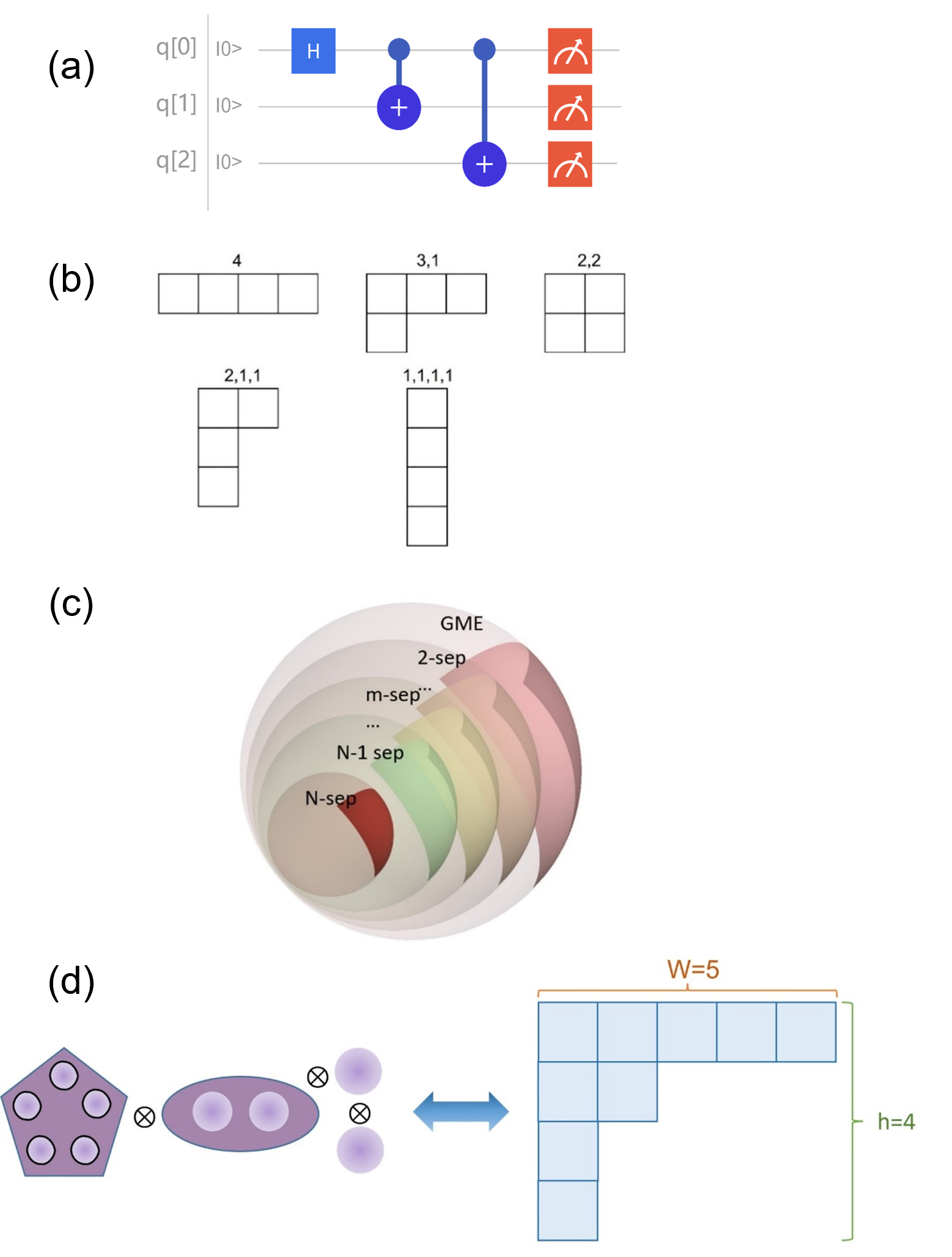}
   
    \caption{Quantum circuit for GHZ state generation and Young diagram representation in the analysis of multi-body entanglement structures. (a) A quantum circuit employed for generating three qubits GHZ state; (b) All possible Young diagrams for a 4-particle system; (c) The separability hierarchy of $N$-qubit states; (d) Characterization of the entanglement structure of a 9-body system using Young diagrams, where it can be divided into $h=4$ subsystems, with an entanglement depth of $w=5$, corresponding to the height $h$ and width $w$ of the Young diagram.}
    \label{fig2}
\end{figure}

The entanglement structure of quantum states refers to entanglement separability and entanglement depth (or entanglement producibility). This can be analogous to the combinatorial problem of N-body system. There are $2^{N-1}$ types of partition, and $(N+1)$ kinds of split methods to decompose an $N$-body system into multiple subsystems $\Lambda=\{\Lambda_1, \Lambda_2, \ldots, \Lambda_k\},\ k\leq N$. Determining the exact analytic equation is very difficult, but recursive and dynamic programming methods can be used to calculate the integer division of $n$ particles. In general, we can use the Young diagram~\cite{ren2021metrological} to directly show the partition. For example, a 4-body system has 8 types of partition and 5 types of split methods. These partitions include 4,$1\otimes3$,$2\otimes2$,$1\otimes1\otimes2$ and $1\otimes1\otimes1\otimes1$. This can be represented by the Young diagram shown in Fig.~\ref{fig2}(b). 

In the Young diagram, each row represents the number of entanglements contained in a subsystem, with the number of entanglements in the subsystems decreasing from top to bottom. Therefore, the width of the top row represents the maximum number of entangled particles, and the number of rows represents the number of separable subsystems.

\textbf{$\boldsymbol{k}$-separable ($\boldsymbol{h}$-inseparable)} is an indicator used to represent the number of separable subsystems in a multi-particle quantum system. A quantum state of $N$ particles is said to be $k$-separable if it can be divided into $k$ groups ($1 \leq k \leq N$) such that there is no entanglement between these groups, although particles within each group may be entangled with each other. Mathematically, a $k$-separable pure state $\rho_{k-sep}$ can be expressed as 

\begin{equation}
\rho_{k-sep}=\rho_{A1}\otimes\rho_{A2}\ldots\otimes\rho_{Ak},
\end{equation}
indicating that the $N$-body system is divided into $k$ separable subsystems. A mixed state is called $k$-separable, if it is a convex combination of $k$-separable pure states. If a quantum state is not $k$-separable, it is considered $h$-inseparable, implying that the system cannot be divided into $h$ separable subsystems.

As shown in Fig.~\ref{fig2}(c), an $N$-qubit state is considered $N$-separable ($N$-sep) if it can be fully decomposed into $N$ independent quantum subsystems, implying that the entire quantum system exhibits no entanglement. On the contrary, if a multi-qubit state cannot be represented as a convex combination of any separable states, it exhibits genuine multipartite entanglement (GME) \cite{toth2005detecting}. GME represents an entanglement structure involving the entire multi-qubit system and cannot be described by decomposing it into smaller subsystems' entanglement structures, indicating that the quantum state exhibits strong quantum correlations among all its subsystems.

\textbf{$\boldsymbol{k}$-producible (Entanglement depth $\boldsymbol{w}$)} is an indicator used to represent the maximum degree of entanglement in the system, which can also be referred to as the entanglement depth $w$. A $k$-producible pure state $\rho_{k-pro}$ can be written as:
\begin{equation}
\rho_{k-pro}=\rho_{A1}\otimes\rho_{A2}\ldots\otimes\rho_{Am}
\label{eq6}
\end{equation}
where $\rho_{Am}$ denotes the state with the maximum number of $k$ entangled particles. If a quantum state is not $k$-producible, its entanglement depth is at least $k+1$, indicating that there are at least $k+1$ particles entangled together in the system.

The larger the value of entanglement depth $w$ or the smaller the value of $k$-separability, the more entanglement that exists in the multipartite system. This concept is intuitively illustrated in Fig.~\ref{fig2}(d).

\section{\label{sec:levelM}Methods}

This section introduces a direct approach to entanglement structure detection based on a multi-view neural network model. The quantum state generation and measurement processes were simulated numerically. These data are encoded as vectors and input into a multi-view neural network which aggregates them into a new representation. The subsequent fully connected neural network \cite{OU20074} classifies various entanglement structures using this representation. The following sections provide detailed descriptions of the dataset preparation, multi-view neural network model, and training process.

\subsection{Dataset preparation}
\begin{figure*}
\centering
\includegraphics[width=0.97\textwidth]{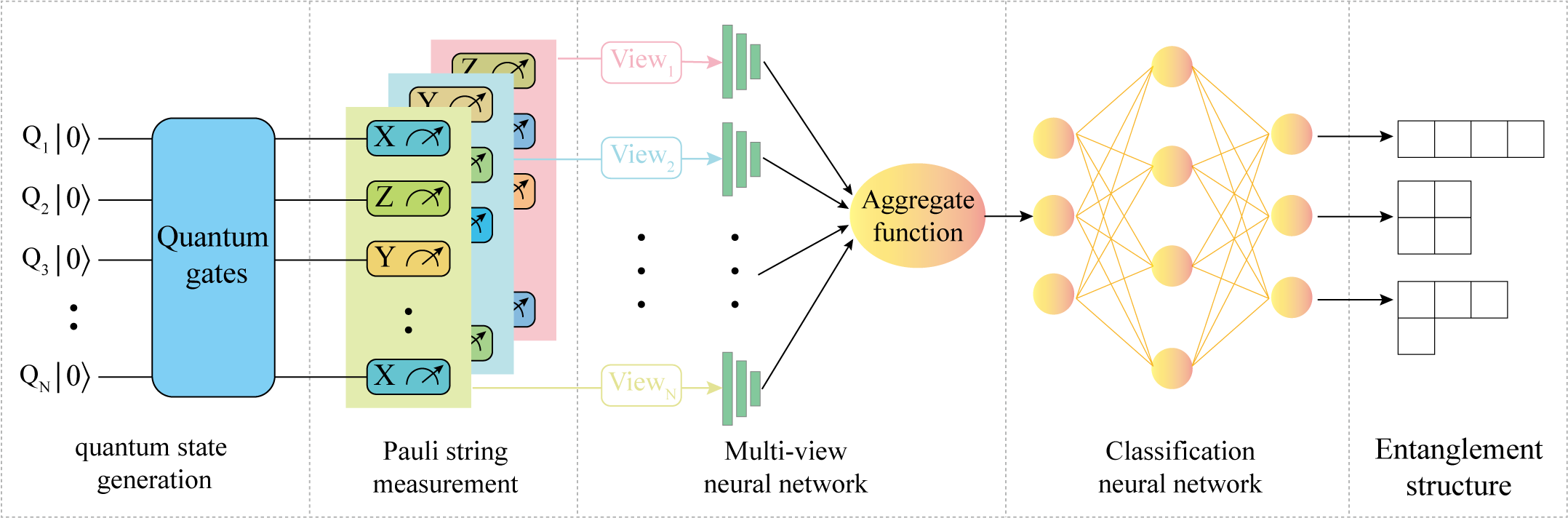}
\caption{Quantum State Generation and Multi-View Neural Network for Entanglement Detection. This figure illustrates the process of quantum state generation, measurement, and subsequent classification using a multi-view neural network. Starting from the left, quantum states for $N$-qubit are initialized in the ground state $\lvert0\rangle$ and processed through a series of quantum gates, generating various entangled states. These states are then measured using Pauli strings, resulting in different views. Each view captures a unique aspect of the quantum state, which is then fed into a multi-view neural network. The network aggregates these views and processes them through fully connected neural network to predict the entanglement structure of the quantum state. This approach enables the model to learn complex interdependencies across different measurements, enhancing the accuracy of quantum state entanglement structure prediction.\label{fig3}}
\end{figure*}

To develop our low-cost quantum entanglement detection method, we generated a dataset of 30,000 quantum states for systems ranging from 4 to 19 qubits using the PennyLane framework ~\cite{bergholm2018pennylane,PennylaneWEB}. This dataset encompasses various entanglement structures, providing a foundation for training and testing our detection algorithm. The data preparation process included three main steps: quantum state generation, label assignment, and measurement simulation, as illustrated in Fig.~\ref{fig3}.

Quantum state generation follows a systematic approach, covering all possible entanglement structures from the GME to completely separable states. First, for each qubit number $N \in {4, \ldots, 19}$, random integer partitions are performed, where each partition represents a combination of subsystems within the $N$-qubit system. Subsequently, based on the partition results, the corresponding quantum states were generated. For example, single-qubit subsystems were generated using random rotation gates, two-qubit subsystems using Bell state circuits, and larger subsystems using GHZ state circuits. This approach allows for comprehensive exploration of all possible subsystem configurations and entanglement depths within each $N$-qubit system.

Next, a precise labeling scheme was employed to accurately label each generated quantum state. The labels correspond to the subsystem composition and entanglement structure. For instance, 'One' denotes a single-qubit subsystem, 'Bell' represents a two-qubit Bell state subsystem, and $GHZ_k$ indicates a $k$-qubit GHZ subsystem. These labels are combined into descriptive strings such as $Bell-One-GHZ_3$, representing a 6 qubits system partition with a Bell state, a single qubit, and a GHZ state. These labels accurately reflect the multifaceted nature of entanglement in complex systems, thereby providing a clear target for our detection algorithm.

To ensure dataset diversity and computational feasibility, we generated all possible structures for systems with $N \leq 8$ qubits. For systems with $N > 8$, we limited the number to 100 structures per qubit count, which were selected randomly based on integer partition results.

Finally, we simulated experimental measurements to mimic data collection on real quantum computers. For each quantum state $\rho$, a set of global measurement operators is performed as follows:

\begin{equation}
\hat{O}=\sigma_{k_1}\otimes\sigma_{k_2}\otimes\ldots\otimes\sigma_{k_N}
\label{eq}
\end{equation}
where Pauli matrices $\sigma_{k}\in\{X, Y, Z\}$. The length of the string corresponds to N qubits. This measurement scheme can capture multi-qubit correlations and global properties of the quantum state. Here, global measurements were chosen because they can comprehensively capture the non-local correlations inherent in the quantum states.
Where Pauli matrices $\sigma_{k} \in {X, Y, Z}$. Global measurements comprehensively capture the non-local correlations inherent in quantum states.

As the system size increases, the number of measurement operators also increases. For systems with 4 to 8 qubits, we simulated all possible global Pauli operator combinations. For larger systems, we employed a stratified sampling approach to select global measurement operators, aiming to retain as much information as possible while reducing computational costs. For 9 and 10 qubits systems, we selected at least 500 different global measurement operators, whereas for systems with more than 10 qubits, we used a minimum of 1000 different global Pauli combinations.

To preserve crucial information, we specifically included global $X$ measurements ($X \otimes X \otimes \ldots \otimes X$), $Y$ measurements ($Y \otimes Y \otimes \ldots \otimes Y$), and $Z$ measurements ($Z \otimes Z \otimes \ldots \otimes Z$), as these measurement operators provide essential information about the quantum state along the principal axes of the Bloch sphere.

To sum up, the final dataset comprised 30,000 unique quantum states, each associated with a specific entanglement structure label, and the simulated outcomes of various global measurement operators. Before being fed into the multi-view neural network model, these data were preprocessed, involving encoding Pauli strings as numerical vectors and converting state descriptions to numerical labels. This encoding facilitates the direct processing of quantum information by neural networks while preserving the structural characteristics of the original data.

\subsection{Multi-view neural network model}

In this section, we introduce the proposed quantum state classification approach based on multi-view learning \cite{wang2015deep}. 
This approach is inspired by the fact that quantum state characterization consists of different measurement operators, each providing a unique perspective on the quantum state. By integrating these diverse viewpoints, we can construct a new representation of quantum states that is suitable for entanglement structure detection. Moreover, the multi-view learning approach is well suited to the complexity and uncertainty of quantum systems. It allows the simultaneous observation and analysis of quantum states from multiple angles, resonating with the principle of complementarity in quantum mechanics. This method can capture subtle features that might be overlooked by single-measurement approaches, thereby improving the overall classification performance.

The core architecture of our model comprises a multi-view neural network and classification neural network as illustrated in Fig.~\ref{fig3}. The multi-view neural network and aggregation module, implemented as a neural network, processes inputs from multiple views, including encoded Pauli string features and probability distributions. 
This module fuses information from multiple views into a unified representation through two fully connected layers using 
ReLU activation \cite{hahnloser2000digital}.
This fusion mechanism enables the model to learn interrelationships and complementary information across different perspectives.

The quantum state classifier is built by adding a fully connected neural network upon this, employing dropout (rate = 0.5) for regularization. The dropout layer (rate=0.5) is employed for regularization to prevent overfitting and improve the generalization capability of the model.

\subsection{Training process}

Our training process utilizes a composite loss function $\mathcal{L}_{\text{total}}$ defined as:
\begin{equation}
\mathcal{L}_{\text{total}} = \mathcal{L}_{\text{CE}} + \lambda \mathcal{L}_{\text{cont}}
\label{eq:total_loss}
\end{equation}
where $\mathcal{L}_{\text{CE}}$ is the cross-entropy loss \cite{zhang2018generalized}, $\mathcal{L}_{\text{cont}}$ is the contrastive loss \cite{khosla2020supervised}, and $\lambda$ is the balancing parameter. This composite loss function serves a dual purpose: the cross-entropy term ensures accurate classification, whereas the contrastive term enhances the model's ability to learn discriminative features.

The cross-entropy loss $\mathcal{L}_{\text{CE}}$ is given by:
\begin{equation}
\mathcal{L}_{\text{CE}} = -\sum_{i=1}^{N} y_i \log(p_i)
\label{eq:cross_entropy}
\end{equation}
where $y_i$ is the one-hot encoded true label and $p_i$ is the model's predicted probability for the $i$-th class. For the contrastive loss $\mathcal{L}_{\text{cont}}$, we first define the cosine similarity function \cite{nguyen2010cosine} between two vectors $\mathbf{x}$ and $\mathbf{y}$ as:
\begin{equation}
\text{cos\_sim}(\mathbf{x}, \mathbf{y}) = \frac{\mathbf{x} \cdot \mathbf{y}}{\|\mathbf{x}\| \|\mathbf{y}\| + \epsilon}
\label{eq:cosine_similarity}
\end{equation}
where $\epsilon = 10^{-8}$ is added for numerical stability. Then, the contrastive loss $\mathcal{L}_{\text{cont}}$ is computed as:
\begin{equation}
\mathcal{L}_{\text{cont}} = \max(0, m - \text{cos\_sim}(\mathbf{a}, \mathbf{p}) + \text{cos\_sim}(\mathbf{a}, \mathbf{n}))
\label{eq:contrastive_loss}
\end{equation}
where $\mathbf{a}$, $\mathbf{p}$, and $\mathbf{n}$ represent the anchor, positive sample, and negative sample, respectively. $m$ is the margin that is set to 1.0 in our implementation. The balancing parameter $\lambda$ in Eq. (\ref{eq:total_loss}) was determined using hyperparameter optimization. A grid search over $\lambda \in [0.001, 1.0]$ was performed with a step size of 0.001 to evaluate the model's performance on a validation set. The optimal value was found to be $\lambda = 0.003$, which provided the best trade-off between the classification accuracy and feature representation quality.


We employed the Adam optimizer \cite{kingma2014adam} for training. The initial learning rate was set to 0.001. To prevent gradient explosion, gradient clipping was applied to limit the gradient norm to 1.0. To ensure stable and efficient training, we use a learning rate scheduler to adjust the learning rate based on validation performance. Training was continued for 50 epochs. After each epoch, the performance was evaluated for the validation set. This iterative process allowed us to monitor learning progress and prevent overfitting. After training, we evaluated its performance on an independent test set to gain a comprehensive understanding of its effectiveness across different quantum states.


\section{\label{sec:levelR}Results Analysis}

This section presents a comprehensive analysis of our multi-view neural network model for entanglement structure detection. We begin by discussing the determination of the optimal number of measurements ($K$-value) with sufficient detection accuracy. We then demonstrate the scalability of our approach up to 19 qubits. Finally, we examined the resource efficiency of our method in terms of quantum measurements and classical computational resources, proving its potential for characterizing larger quantum systems.

\subsection{$K$-value Selection Condition}

The number of required quantum measurement operators, denoted by the $K$-value, is critical to the success of our approach. Its determination is mainly based on detection performance. Two indicators, detection accuracy and precision of each structure in a specific particle system were adopted to analyze the performance of various measurement configurations.

\begin{figure}[htbp]
    \centering
    \includegraphics[width=0.49\textwidth]{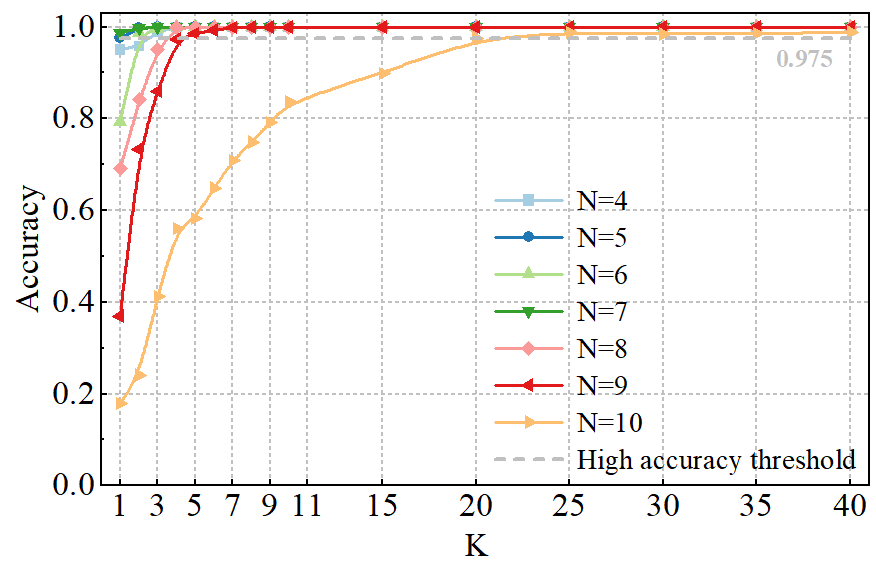}
    \caption{The figure shows the detection accuracy of our method for quantum systems ranging from 4 to 10 qubits (N=4 to 10) as a function of the number of measurement operators (K). The dashed line at 0.975 represents our high accuracy threshold.}
    \label{fig:performance_vs_K}
\end{figure}

The detection accuracy assesses the capability of the proposed method to identify complex entanglement structures. For systems with N $\leq$ 10, evaluated for all possible K-values were evaluated. The results are illustrated in Fig. \ref{fig:performance_vs_K}. To ensure reliable detection, we set a high accuracy threshold of 0.975. The optimal K-value was identified as the minimum number of measurements across this threshold. It can be seen that for smaller systems (N $\leq$ 7), this high accuracy threshold can be achieved with very few measurements (K $\leq$ 3). As the system size increased to N=8 and N=9, more measurements are required, but the number remained manageable. Even for a system with up to 10 qubits, approximately 20 measurements were sufficient to reach the 0.97 accuracy benchmark.

\begin{table}[h]
\caption{Precision for different entanglement structures in a 4 qubits system.}
\centering
\begin{tabular}{|c|c|}
\hline
Entanglement Structure & Precision \\
\hline
GHZ\_4 & 0.984 \\
Bell-Bell & 1.00 \\
One-GHZ\_3  & 0.934 \\
GHZ\_3-One  & 0.995 \\
Bell-One-One & 1.00 \\
One-Bell-One  & 0.974 \\
One-One-Bell  & 0.988 \\
One-One-One-One  & 0.995 \\
\hline
\end{tabular}
\label{tab:4_particle_accuracy}
\end{table}

To further demonstrate the performance of our method, the precision of identifying specific entanglement structures within a quantum system was evaluated. Table \ref{tab:4_particle_accuracy} summarizes the results for a 4 qubits system with K=2. This detailed evaluation shows that our method maintains a high accuracy across various entanglement structures, although the detection accuracy varies slightly among structures.

Based on the above results for systems with N $\leq$ 10, we derived an empirical formula to determine the optimal $K$-value as follows:
\begin{equation}
K = 8.6 \times 10^{-14} \cdot N^{14.31} + 1.82
\label{eq:k}
\end{equation}
The calculated $K$-value should be rounded to ensure that the number of measurements is sufficient to achieve the desired accuracy. The empirical formula shows that the number of required quantum measurements of our method is polynomial rather than exponentially increasing with the qubit number. 
This indicates a remarkable reduction in the number of required quantum measurements, which will be discussed in Section (\ref{tab:resource}). More importantly, the massive size of dataset and the computational cost make it impossible to determine the K-value by exhaustively evaluating the performance of all measurement configurations for large systems. Then, this empirical formula can provide practical guidance for determining the number of measurements required to efficiently and accurately detect entanglement structures in systems larger than 10 qubits. This would be discussed in Section (\ref{tab:large}).



By considering both the detection accuracy and the precision of specific structures in our analysis, we ensured a comprehensive evaluation of our method's performance and thus a reasonable determination of the number of required quantum measurements. This multifaceted approach highlights the efficiency and scalability of our method, paving the way for a more effective characterization of entanglement structures in larger quantum systems.

\subsection{\label{tab:large} Scalability up to 19 Qubits}

\begin{figure}[htbp]
\centering
\includegraphics[width=0.49\textwidth]{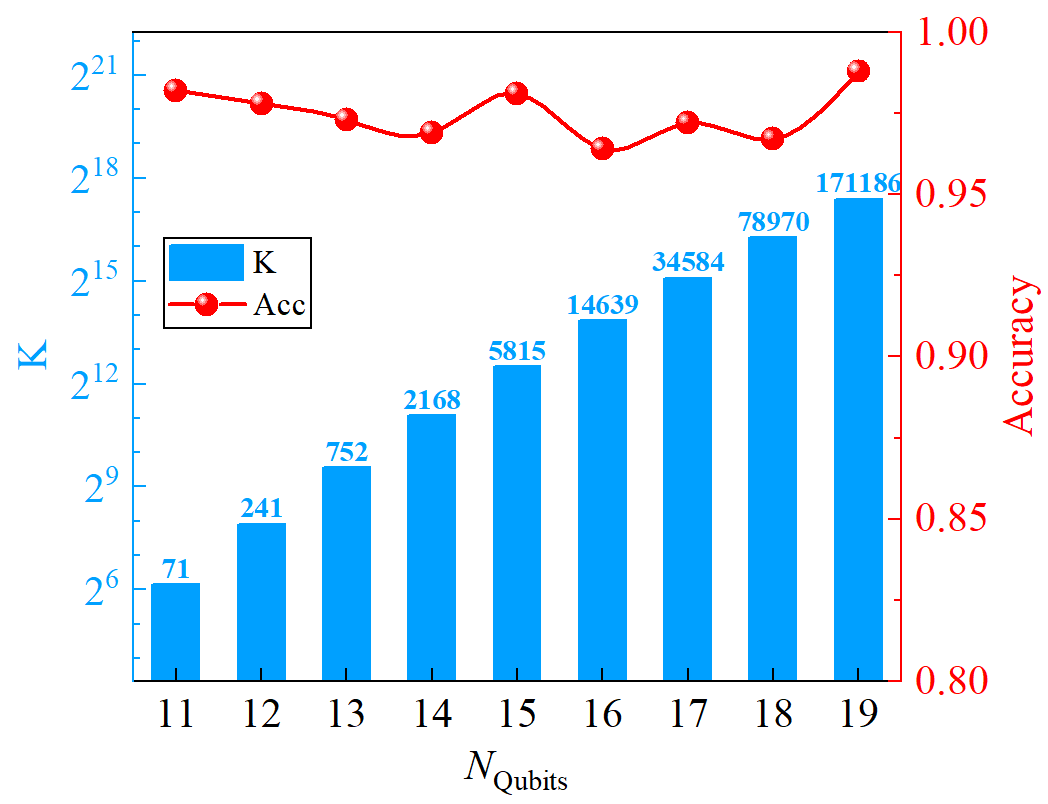}
\caption{Detection Accuracy and Measurement Requirements for Systems with N=11 to 19 Qubits. The bars represent the required measurements, while the red line depicts the detection accuracy as a percentage. Note the consistently high accuracy despite the increasing system complexity.}
\label{fig5}
\end{figure}

In this section, we describe the detection performance of our approach for systems with 11 to 19 qubits. First, detection accuracies were evaluated for various quantum systems. The K-value was determined by the empirical formula in Eq.(\ref{eq:k}). The results are shown in Fig. \ref{fig5}. It is demonstrated that our approach achieves high detection accuracy (exceeding 95\%) across systems with 11 to 19 qubits. This consistent accuracy across varying qubit numbers provided robust support for the validity and scalability of the proposed empirical formula. It should be pointed out that fluctuations in the detection accuracy, particularly at 15 and 19 qubits, might be related to the structural characteristics or the distribution of entangled states in these specific quantum systems. In such cases, the neural network may exhibit a higher sensitivity to certain entanglement structures, leading to improved detection accuracy. Although these fluctuations are noteworthy, they do not affect the overall scalability and robustness of the method, as the detection accuracy remains above 95\% across all tested qubit systems.

To further validate the reliability of the empirical formula, we evaluated the precision of our method for detecting various entanglement structures within a 19 qubits system. 
Table \ref{tab:19_particle_accuracy} provides a detailed breakdown of the detection accuracy of these structures. It is shown that our method maintains remarkably high detection accuracy across diverse entanglement structures within the 19 qubits system. This further confirms that the K-value derived from our empirical formula is not only effective for general system sizes but also reliable for accurately detecting complex entanglement structures in large-scale quantum systems, a crucial capability for advanced quantum information protocols.

\begin{table}[h]
\caption{Precision for different entanglement structures in a 19 qubits system.}
\centering
\begin{tabular}{|p{7cm}|p{1.5cm}|}
\hline
Entanglement Structure & Precision \\
\hline
GHZ\_19 & 0.99 \\
One-GHZ\_18  & 0.987 \\
Bell-Bell-GHZ\_3-GHZ\_3-Bell-GHZ\_6-One & 1.00 \\
Bell-GHZ\_3-One-One-GHZ\_3-GHZ\_9  & 0.997 \\
Bell-Bell-One-One-One-GHZ\_4-GHZ\_3-GHZ\_5  & 0.977 \\
One-One-One-Bell-GHZ\_10-One-Bell-One & 0.986 \\
One-One-One-GHZ\_4-GHZ\_4-Bell-GHZ\_6  & 0.978 \\
One-Bell-Bell-Bell-Bell-One-Bell-GHZ\_7  & 1 \\
GHZ\_3-One-One-One-One-Bell-One-GHZ\_4-One-GHZ\_4  & 0.99 \\
GHZ\_3-One-One-One-GHZ\_12-One  & 0.99 \\
\hline
\end{tabular}
\label{tab:19_particle_accuracy}
\end{table}

Both the detection accuracy and precision of specific structures provide compelling evidence for the reliability and practicality of our empirical formula. This validation underscores the effectiveness of our method in significantly reducing the number of required measurements while maintaining a high accuracy.

\subsection{\label{tab:resource} Resource efficiency}


In this section, we compare the number of required quantum measurements of our model with that of QST while comparing the needed classical computational resource of our model with that of other data-driven methods based on the density matrix.

As mentioned above, the conventional QST approach requires $4^n-1$ independent measurements for the $n$-qubit state characterization. Although this method provides complete information about the quantum state, it becomes impractical for systems with more than a few qubits owing to the exponential growth in measurement requirements. To address this limitation, advanced techniques such as compressed sensing QST (CS-QST)~\cite{rodionov2014compressed,gebhart2023learning,gross2010quantum}, neural network QST(NN-QST)~\cite{PRXQuantum.2.020348} and adaptive measurement QST (AM-QST)~\cite{fei2017adaptive,quek2021adaptive} have been developed.

\begin{figure}[htbp]
\centering
\includegraphics[width=0.49\textwidth]{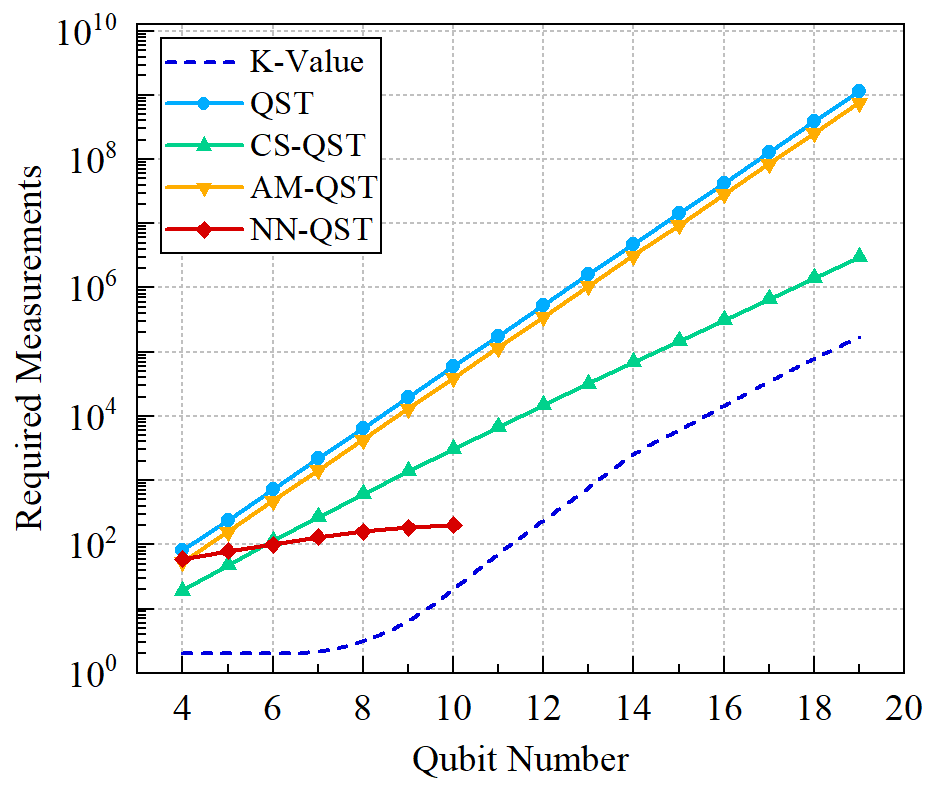}
\caption{Comparison of required quantum measurements as a function of qubit number for different quantum state tomography methods. The plotted methods include conventional QST, CS-QST~\cite{gross2010quantum}, AM-QST~\cite{PhysRevApplied.20.064007}, NN-QST~\cite{PRXQuantum.2.020348}, and our proposed method represented by the K-value curve.}

\label{fig6}
\end{figure}

Fig.\ref{fig6} illustrates the comparison results of the proposed approach with the established QST techniques. For a quantum state with rank $r$, the measurement complexity of CS-QST is approximately $M \propto r \cdot 2^n \log(2^n)$~\cite{gross2010quantum}. NN-QST methods have been proven to be highly effective in performing QST for small to medium-sized systems. The detailed results can be found in Ref.~\cite{PRXQuantum.2.020348}. Recent work on AM-QST expects to reduce the number of required measurements by approximately 33.74\%~\cite{PhysRevApplied.20.064007}. It can be seen that our multi-view learning model significantly reduces the number of required quantum measurements since our method maintains polynomial growth in the number of measurements as the system size grows while other methods exhibit exponential growth. This advantage becomes more pronounced for larger systems, suggesting superior scalability for complex quantum states. To better demonstrate this, we calculated the reduction factor in the measurements compared to the conventional QST. For a 6 qubits system, our method achieves a reduction factor of approximately 364×, which increases to approximately 2,952× for a 10 qubits system and 6,789× for a 19 qubits system. These reduction factors highlight the dramatic decrease in required measurements achieved by our approach, with improvements of three to four orders of magnitude.

\begin{table*}[t]
\centering
\begin{longtable}{|c|c|c|c|c|}
\caption{Comparison of memory usage between our method and density matrix approaches.}
\label{tab:memory_comparison} \\
\hline
\textbf{Qubit number} 
& \makecell{\textbf{Input Size} \\ \textbf{(Our Method)}} 
& \makecell{\textbf{Memory Usage} \\ \textbf{(Our Method)}} 
& \makecell{\textbf{Input Size} \\ \textbf{(Density Matrix)}} 
& \makecell{\textbf{Memory Usage} \\ \textbf{(Density Matrix)}} \\

\hline
4  & 28    & 1.31 MB  & 256       & 1.75 MB  \\
5  & 47    & 1.35 MB  & 1,024      & 3.26 MB  \\
6  & 82    & 1.41 MB  & 4,096      & 9.27 MB  \\
7  & 149   & 1.55 MB  & 16,384     & 33.32 MB \\
8  & 280   & 1.80 MB  & 65,536     & 129.50 MB\\
9  & 539   & 2.31 MB  & 262,144    & 514.25 MB\\
10 & 1,054  & 3.32 MB  & 1,048,576   & 2.01 GB  \\
11 & 2,081  & 5.33 MB  & 4,194,304   & 8.02 GB  \\
12 & 4,132  & 9.34 MB  & 16,777,216  & 32.06 GB \\
13 & 8,231  & 17.36 MB & 67,108,864  & 128.25 GB\\
14 & 16,426 & 33.40 MB & 268,435,456 & 513.00 GB\\
15 & 32,813 & 65.47 MB & 1,073,741,824& 2.00 TB  \\
16 & 65,584 & 129.60 MB& 4,294,967,296& 8.02 TB  \\
17 & 131,123& 257.85 MB& 17,179,869,184& 32.06 TB \\
18 & 262,198& 514.36 MB& 68,719,476,736& 128.25 TB\\
19 & 524,345& 1.00 GB  & 274,877,906,944& 513.00 TB\\
\hline

\end{longtable}
\end{table*}

Beyond the measurement requirements, our method also possesses significant efficiency in terms of classical computational resources, particularly in terms of memory usage. Table \ref{tab:memory_comparison} compares the memory requirements of our approach with those of the same classification neural network based on density matrix. In our approach, the input for each iteration consists of the Pauli string encoding (a $3 \times N$ array) and measurement results (a probability distribution of length $2^N$). In contrast, the dimensions of the density matrix were $2^N \times 2^N$. Inputting only one row of the density matrix into the neural network is practically meaningless, as it fails to capture the complete quantum state information. The results show the proposed method consistently requires significantly less memory. As the system size increases, this advantage becomes even more pronounced. For a 19 qubits system, the proposed method uses only 1 GB of memory. This reduction in memory usage, coupled with a reduction in measurement requirements, results in a substantial improvement in the computational efficiency.

\section{\label{sec:level4} Conclusion}

In this study, we developed a multi-view neural network approach for efficient multipartite entanglement structure detection in large-scale quantum systems. This approach breaks the general routine of entanglement detection which performs state tomography and then applies a detection method to the reconstructed density matrix. A multi-view neural network with finite global measurement outcomes as input generates an alternative representation. In view of this, our approach has three important practical properties.

First, it has universality. The practical difficulties of certifying high-dimensional entanglement have motivated some preconditions to simplify the problem \cite{martin2017quantifying,chang2021648}. Here, we make no such requirement. Furthermore, our method can be applied to classify various entanglement structures from GME to completely separable states. This makes it more powerful than EW which can only distinguish entangled and separable states directly by sophisticatedly designed measurement operators \cite{friis2019entanglement,erhard2020advances}.

Secondly, it requires little resource cost. An empirical formula of K-value is derived based on the data of systems with N $\leq$ 10. Numerical simulations with 11- to 19 qubits systems validate its reliability. This empirical formula indicates the scaling of the number of required measurements is significantly reduced into a polynomial increase with qubit number, making viable detection of high-dimensional entanglement structures possible. Besides, thanks to the generated representation, our method requires substantially less memory. This implies more computational efficiency and less time consumption.

Thirdly, it permits the experimentalist to select the number of measurements used for entanglement structure detection. Notably, in our work, the K-value was identified as the minimal number of measurements across a high accuracy threshold of 97\%. This leaves space for adjustment. For instance, the predicted K-value for the system with 19 qubits had a detection accuracy of 97\%. 
If we reduce the number of measurements to 160,000, the detection accuracy would decrease to 95.35\%.

\begin{acknowledgments}
The authors would like to express their sincere gratitude to Dr. Zhen Chen for his invaluable support and insightful discussions that greatly contributed to this work. This work is supported by the National Natural Science Foundation of China under Grants No.61975005, Beijing Academy of Quantum Information Science under Grants No.Y18G28, the Industry University Research Cooperation Fund of the Eighth Research Institute of China Aerospace Science and Technology Corporation No.SAST2023-030, and the Fundamental Research Funds for the Central Universities under Grants No.YWF-22-L-938.
\end{acknowledgments}

\nocite{*}

\bibliography{apssamp}

\begin{thebibliography}{55}%
\makeatletter
\providecommand \@ifxundefined [1]{%
 \@ifx{#1\undefined}
}%
\providecommand \@ifnum [1]{%
 \ifnum #1\expandafter \@firstoftwo
 \else \expandafter \@secondoftwo
 \fi
}%
\providecommand \@ifx [1]{%
 \ifx #1\expandafter \@firstoftwo
 \else \expandafter \@secondoftwo
 \fi
}%
\providecommand \natexlab [1]{#1}%
\providecommand \enquote  [1]{``#1''}%
\providecommand \bibnamefont  [1]{#1}%
\providecommand \bibfnamefont [1]{#1}%
\providecommand \citenamefont [1]{#1}%
\providecommand \href@noop [0]{\@secondoftwo}%
\providecommand \href [0]{\begingroup \@sanitize@url \@href}%
\providecommand \@href[1]{\@@startlink{#1}\@@href}%
\providecommand \@@href[1]{\endgroup#1\@@endlink}%
\providecommand \@sanitize@url [0]{\catcode `\\12\catcode `\$12\catcode `\&12\catcode `\#12\catcode `\^12\catcode `\_12\catcode `\%12\relax}%
\providecommand \@@startlink[1]{}%
\providecommand \@@endlink[0]{}%
\providecommand \url  [0]{\begingroup\@sanitize@url \@url }%
\providecommand \@url [1]{\endgroup\@href {#1}{\urlprefix }}%
\providecommand \urlprefix  [0]{URL }%
\providecommand \Eprint [0]{\href }%
\providecommand \doibase [0]{https://doi.org/}%
\providecommand \selectlanguage [0]{\@gobble}%
\providecommand \bibinfo  [0]{\@secondoftwo}%
\providecommand \bibfield  [0]{\@secondoftwo}%
\providecommand \translation [1]{[#1]}%
\providecommand \BibitemOpen [0]{}%
\providecommand \bibitemStop [0]{}%
\providecommand \bibitemNoStop [0]{.\EOS\space}%
\providecommand \EOS [0]{\spacefactor3000\relax}%
\providecommand \BibitemShut  [1]{\csname bibitem#1\endcsname}%
\let\auto@bib@innerbib\@empty
\bibitem [{\citenamefont {Einstein}\ \emph {et~al.}(1935)\citenamefont {Einstein}, \citenamefont {Podolsky},\ and\ \citenamefont {Rosen}}]{einstein1935can}%
  \BibitemOpen
  \bibfield  {author} {\bibinfo {author} {\bibfnamefont {A.}~\bibnamefont {Einstein}}, \bibinfo {author} {\bibfnamefont {B.}~\bibnamefont {Podolsky}},\ and\ \bibinfo {author} {\bibfnamefont {N.}~\bibnamefont {Rosen}},\ }\bibfield  {title} {\bibinfo {title} {Can quantum-mechanical description of physical reality be considered complete?},\ }\href@noop {} {\bibfield  {journal} {\bibinfo  {journal} {Physical review}\ }\textbf {\bibinfo {volume} {47}},\ \bibinfo {pages} {777} (\bibinfo {year} {1935})}\BibitemShut {NoStop}%
\bibitem [{\citenamefont {Schr{\"o}dinger}(1935)}]{schrodinger1935discussion}%
  \BibitemOpen
  \bibfield  {author} {\bibinfo {author} {\bibfnamefont {E.}~\bibnamefont {Schr{\"o}dinger}},\ }\bibfield  {title} {\bibinfo {title} {Discussion of probability relations between separated systems},\ }in\ \href@noop {} {\emph {\bibinfo {booktitle} {Mathematical Proceedings of the Cambridge Philosophical Society}}},\ Vol.~\bibinfo {volume} {31}\ (\bibinfo {organization} {Cambridge University Press},\ \bibinfo {year} {1935})\ pp.\ \bibinfo {pages} {555--563}\BibitemShut {NoStop}%
\bibitem [{\citenamefont {Chitambar}\ and\ \citenamefont {Gour}(2019)}]{chitambar2019quantum}%
  \BibitemOpen
  \bibfield  {author} {\bibinfo {author} {\bibfnamefont {E.}~\bibnamefont {Chitambar}}\ and\ \bibinfo {author} {\bibfnamefont {G.}~\bibnamefont {Gour}},\ }\bibfield  {title} {\bibinfo {title} {Quantum resource theories},\ }\href@noop {} {\bibfield  {journal} {\bibinfo  {journal} {Reviews of modern physics}\ }\textbf {\bibinfo {volume} {91}},\ \bibinfo {pages} {025001} (\bibinfo {year} {2019})}\BibitemShut {NoStop}%
\bibitem [{\citenamefont {Ehlers}\ \emph {et~al.}(2015)\citenamefont {Ehlers}, \citenamefont {S\'olyom}, \citenamefont {Legeza},\ and\ \citenamefont {Noack}}]{PhysRevB.92.235116}%
  \BibitemOpen
  \bibfield  {author} {\bibinfo {author} {\bibfnamefont {G.}~\bibnamefont {Ehlers}}, \bibinfo {author} {\bibfnamefont {J.}~\bibnamefont {S\'olyom}}, \bibinfo {author} {\bibfnamefont {O.}~\bibnamefont {Legeza}},\ and\ \bibinfo {author} {\bibfnamefont {R.~M.}\ \bibnamefont {Noack}},\ }\bibfield  {title} {\bibinfo {title} {Entanglement structure of the hubbard model in momentum space},\ }\href {https://doi.org/10.1103/PhysRevB.92.235116} {\bibfield  {journal} {\bibinfo  {journal} {Phys. Rev. B}\ }\textbf {\bibinfo {volume} {92}},\ \bibinfo {pages} {235116} (\bibinfo {year} {2015})}\BibitemShut {NoStop}%
\bibitem [{\citenamefont {Alba}(2018)}]{PhysRevB.97.245135}%
  \BibitemOpen
  \bibfield  {author} {\bibinfo {author} {\bibfnamefont {V.}~\bibnamefont {Alba}},\ }\bibfield  {title} {\bibinfo {title} {Entanglement and quantum transport in integrable systems},\ }\href {https://doi.org/10.1103/PhysRevB.97.245135} {\bibfield  {journal} {\bibinfo  {journal} {Phys. Rev. B}\ }\textbf {\bibinfo {volume} {97}},\ \bibinfo {pages} {245135} (\bibinfo {year} {2018})}\BibitemShut {NoStop}%
\bibitem [{\citenamefont {Laflorencie}(2016)}]{laflorencie2016quantum}%
  \BibitemOpen
  \bibfield  {author} {\bibinfo {author} {\bibfnamefont {N.}~\bibnamefont {Laflorencie}},\ }\bibfield  {title} {\bibinfo {title} {Quantum entanglement in condensed matter systems},\ }\href@noop {} {\bibfield  {journal} {\bibinfo  {journal} {Physics Reports}\ }\textbf {\bibinfo {volume} {646}},\ \bibinfo {pages} {1} (\bibinfo {year} {2016})}\BibitemShut {NoStop}%
\bibitem [{\citenamefont {Feynman}(1982)}]{feynman1982simulating}%
  \BibitemOpen
  \bibfield  {author} {\bibinfo {author} {\bibfnamefont {R.~P.}\ \bibnamefont {Feynman}},\ }\bibfield  {title} {\bibinfo {title} {Simulating physics with computers},\ }\href@noop {} {\bibfield  {journal} {\bibinfo  {journal} {International Journal of Theoretical Physics}\ }\textbf {\bibinfo {volume} {21}},\ \bibinfo {pages} {467} (\bibinfo {year} {1982})}\BibitemShut {NoStop}%
\bibitem [{\citenamefont {Zhang}\ \emph {et~al.}(2024)\citenamefont {Zhang}, \citenamefont {Qin}, \citenamefont {Zhou}, \citenamefont {Li}, \citenamefont {Du},\ and\ \citenamefont {Xiao}}]{zhang2024single}%
  \BibitemOpen
  \bibfield  {author} {\bibinfo {author} {\bibfnamefont {S.}~\bibnamefont {Zhang}}, \bibinfo {author} {\bibfnamefont {Z.}~\bibnamefont {Qin}}, \bibinfo {author} {\bibfnamefont {Y.}~\bibnamefont {Zhou}}, \bibinfo {author} {\bibfnamefont {R.}~\bibnamefont {Li}}, \bibinfo {author} {\bibfnamefont {C.}~\bibnamefont {Du}},\ and\ \bibinfo {author} {\bibfnamefont {Z.}~\bibnamefont {Xiao}},\ }\bibfield  {title} {\bibinfo {title} {Single entanglement connection architecture between multi-layer bipartite hardware efficient ansatz},\ }\href {https://doi.org/10.1088/1367-2630/ad64fb} {\bibfield  {journal} {\bibinfo  {journal} {New Journal of Physics}\ }\textbf {\bibinfo {volume} {26}},\ \bibinfo {pages} {073042} (\bibinfo {year} {2024})}\BibitemShut {NoStop}%
\bibitem [{\citenamefont {Qin}\ \emph {et~al.}(2024)\citenamefont {Qin}, \citenamefont {Li}, \citenamefont {Zhou}, \citenamefont {Zhang}, \citenamefont {Li}, \citenamefont {Du},\ and\ \citenamefont {Xiao}}]{qin2024applicability}%
  \BibitemOpen
  \bibfield  {author} {\bibinfo {author} {\bibfnamefont {Z.}~\bibnamefont {Qin}}, \bibinfo {author} {\bibfnamefont {X.}~\bibnamefont {Li}}, \bibinfo {author} {\bibfnamefont {Y.}~\bibnamefont {Zhou}}, \bibinfo {author} {\bibfnamefont {S.}~\bibnamefont {Zhang}}, \bibinfo {author} {\bibfnamefont {R.}~\bibnamefont {Li}}, \bibinfo {author} {\bibfnamefont {C.}~\bibnamefont {Du}},\ and\ \bibinfo {author} {\bibfnamefont {Z.}~\bibnamefont {Xiao}},\ }\bibfield  {title} {\bibinfo {title} {Applicability of measurement-based quantum computation towards physically-driven variational quantum eigensolver},\ }\href {https://doi.org/10.1088/1367-2630/ad634a} {\bibfield  {journal} {\bibinfo  {journal} {New Journal of Physics}\ }\textbf {\bibinfo {volume} {26}},\ \bibinfo {pages} {073040} (\bibinfo {year} {2024})}\BibitemShut {NoStop}%
\bibitem [{\citenamefont {Alber}\ \emph {et~al.}(2001)\citenamefont {Alber}, \citenamefont {Beth}, \citenamefont {Horodecki}, \citenamefont {Horodecki}, \citenamefont {Horodecki}, \citenamefont {R{\"o}tteler}, \citenamefont {Weinfurter}, \citenamefont {Werner}, \citenamefont {Zeilinger}, \citenamefont {Horodecki} \emph {et~al.}}]{alber2001mixed}%
  \BibitemOpen
  \bibfield  {author} {\bibinfo {author} {\bibfnamefont {G.}~\bibnamefont {Alber}}, \bibinfo {author} {\bibfnamefont {T.}~\bibnamefont {Beth}}, \bibinfo {author} {\bibfnamefont {M.}~\bibnamefont {Horodecki}}, \bibinfo {author} {\bibfnamefont {P.}~\bibnamefont {Horodecki}}, \bibinfo {author} {\bibfnamefont {R.}~\bibnamefont {Horodecki}}, \bibinfo {author} {\bibfnamefont {M.}~\bibnamefont {R{\"o}tteler}}, \bibinfo {author} {\bibfnamefont {H.}~\bibnamefont {Weinfurter}}, \bibinfo {author} {\bibfnamefont {R.}~\bibnamefont {Werner}}, \bibinfo {author} {\bibfnamefont {A.}~\bibnamefont {Zeilinger}}, \bibinfo {author} {\bibfnamefont {M.}~\bibnamefont {Horodecki}}, \emph {et~al.},\ }\bibfield  {title} {\bibinfo {title} {Mixed-state entanglement and quantum communication},\ }\href@noop {} {\bibfield  {journal} {\bibinfo  {journal} {Quantum information: An introduction to basic theoretical concepts and experiments}\ ,\ \bibinfo {pages} {151}} (\bibinfo {year} {2001})}\BibitemShut {NoStop}%
\bibitem [{\citenamefont {Ursin}\ \emph {et~al.}(2007)\citenamefont {Ursin}, \citenamefont {Tiefenbacher}, \citenamefont {Schmitt-Manderbach}, \citenamefont {Weier}, \citenamefont {Scheidl}, \citenamefont {Lindenthal}, \citenamefont {Blauensteiner}, \citenamefont {Jennewein}, \citenamefont {Perdigues}, \citenamefont {Trojek} \emph {et~al.}}]{ursin2007entanglement}%
  \BibitemOpen
  \bibfield  {author} {\bibinfo {author} {\bibfnamefont {R.}~\bibnamefont {Ursin}}, \bibinfo {author} {\bibfnamefont {F.}~\bibnamefont {Tiefenbacher}}, \bibinfo {author} {\bibfnamefont {T.}~\bibnamefont {Schmitt-Manderbach}}, \bibinfo {author} {\bibfnamefont {H.}~\bibnamefont {Weier}}, \bibinfo {author} {\bibfnamefont {T.}~\bibnamefont {Scheidl}}, \bibinfo {author} {\bibfnamefont {M.}~\bibnamefont {Lindenthal}}, \bibinfo {author} {\bibfnamefont {B.}~\bibnamefont {Blauensteiner}}, \bibinfo {author} {\bibfnamefont {T.}~\bibnamefont {Jennewein}}, \bibinfo {author} {\bibfnamefont {J.}~\bibnamefont {Perdigues}}, \bibinfo {author} {\bibfnamefont {P.}~\bibnamefont {Trojek}}, \emph {et~al.},\ }\bibfield  {title} {\bibinfo {title} {Entanglement-based quantum communication over 144 km},\ }\href@noop {} {\bibfield  {journal} {\bibinfo  {journal} {Nature physics}\ }\textbf {\bibinfo {volume} {3}},\ \bibinfo {pages} {481} (\bibinfo {year} {2007})}\BibitemShut {NoStop}%
\bibitem [{\citenamefont {Pan}\ \emph {et~al.}(2001)\citenamefont {Pan}, \citenamefont {Simon}, \citenamefont {Brukner},\ and\ \citenamefont {Zeilinger}}]{pan2001entanglement}%
  \BibitemOpen
  \bibfield  {author} {\bibinfo {author} {\bibfnamefont {J.-W.}\ \bibnamefont {Pan}}, \bibinfo {author} {\bibfnamefont {C.}~\bibnamefont {Simon}}, \bibinfo {author} {\bibfnamefont {{\v{C}}.}~\bibnamefont {Brukner}},\ and\ \bibinfo {author} {\bibfnamefont {A.}~\bibnamefont {Zeilinger}},\ }\bibfield  {title} {\bibinfo {title} {Entanglement purification for quantum communication},\ }\href@noop {} {\bibfield  {journal} {\bibinfo  {journal} {Nature}\ }\textbf {\bibinfo {volume} {410}},\ \bibinfo {pages} {1067} (\bibinfo {year} {2001})}\BibitemShut {NoStop}%
\bibitem [{\citenamefont {Pezze}\ \emph {et~al.}(2018)\citenamefont {Pezze}, \citenamefont {Smerzi}, \citenamefont {Oberthaler}, \citenamefont {Schmied},\ and\ \citenamefont {Treutlein}}]{pezze2018quantum}%
  \BibitemOpen
  \bibfield  {author} {\bibinfo {author} {\bibfnamefont {L.}~\bibnamefont {Pezze}}, \bibinfo {author} {\bibfnamefont {A.}~\bibnamefont {Smerzi}}, \bibinfo {author} {\bibfnamefont {M.~K.}\ \bibnamefont {Oberthaler}}, \bibinfo {author} {\bibfnamefont {R.}~\bibnamefont {Schmied}},\ and\ \bibinfo {author} {\bibfnamefont {P.}~\bibnamefont {Treutlein}},\ }\bibfield  {title} {\bibinfo {title} {Quantum metrology with nonclassical states of atomic ensembles},\ }\href@noop {} {\bibfield  {journal} {\bibinfo  {journal} {Reviews of Modern Physics}\ }\textbf {\bibinfo {volume} {90}},\ \bibinfo {pages} {035005} (\bibinfo {year} {2018})}\BibitemShut {NoStop}%
\bibitem [{\citenamefont {Ren}\ \emph {et~al.}(2021)\citenamefont {Ren}, \citenamefont {Li}, \citenamefont {Smerzi},\ and\ \citenamefont {Gessner}}]{ren2021metrological}%
  \BibitemOpen
  \bibfield  {author} {\bibinfo {author} {\bibfnamefont {Z.}~\bibnamefont {Ren}}, \bibinfo {author} {\bibfnamefont {W.}~\bibnamefont {Li}}, \bibinfo {author} {\bibfnamefont {A.}~\bibnamefont {Smerzi}},\ and\ \bibinfo {author} {\bibfnamefont {M.}~\bibnamefont {Gessner}},\ }\bibfield  {title} {\bibinfo {title} {Metrological detection of multipartite entanglement from young diagrams},\ }\href@noop {} {\bibfield  {journal} {\bibinfo  {journal} {Physical Review Letters}\ }\textbf {\bibinfo {volume} {126}},\ \bibinfo {pages} {080502} (\bibinfo {year} {2021})}\BibitemShut {NoStop}%
\bibitem [{\citenamefont {Gross}\ \emph {et~al.}(2009)\citenamefont {Gross}, \citenamefont {Flammia},\ and\ \citenamefont {Eisert}}]{PhysRevLett.102.190501}%
  \BibitemOpen
  \bibfield  {author} {\bibinfo {author} {\bibfnamefont {D.}~\bibnamefont {Gross}}, \bibinfo {author} {\bibfnamefont {S.~T.}\ \bibnamefont {Flammia}},\ and\ \bibinfo {author} {\bibfnamefont {J.}~\bibnamefont {Eisert}},\ }\bibfield  {title} {\bibinfo {title} {Most quantum states are too entangled to be useful as computational resources},\ }\href {https://doi.org/10.1103/PhysRevLett.102.190501} {\bibfield  {journal} {\bibinfo  {journal} {Phys. Rev. Lett.}\ }\textbf {\bibinfo {volume} {102}},\ \bibinfo {pages} {190501} (\bibinfo {year} {2009})}\BibitemShut {NoStop}%
\bibitem [{\citenamefont {Dür}\ and\ \citenamefont {Briegel}(2007)}]{Dür_2007}%
  \BibitemOpen
  \bibfield  {author} {\bibinfo {author} {\bibfnamefont {W.}~\bibnamefont {Dür}}\ and\ \bibinfo {author} {\bibfnamefont {H.~J.}\ \bibnamefont {Briegel}},\ }\bibfield  {title} {\bibinfo {title} {Entanglement purification and quantum error correction},\ }\href {https://doi.org/10.1088/0034-4885/70/8/R03} {\bibfield  {journal} {\bibinfo  {journal} {Reports on Progress in Physics}\ }\textbf {\bibinfo {volume} {70}},\ \bibinfo {pages} {1381} (\bibinfo {year} {2007})}\BibitemShut {NoStop}%
\bibitem [{\citenamefont {Horodecki}\ \emph {et~al.}(2009)\citenamefont {Horodecki}, \citenamefont {Horodecki}, \citenamefont {Horodecki},\ and\ \citenamefont {Horodecki}}]{RevModPhys.81.865}%
  \BibitemOpen
  \bibfield  {author} {\bibinfo {author} {\bibfnamefont {R.}~\bibnamefont {Horodecki}}, \bibinfo {author} {\bibfnamefont {P.}~\bibnamefont {Horodecki}}, \bibinfo {author} {\bibfnamefont {M.}~\bibnamefont {Horodecki}},\ and\ \bibinfo {author} {\bibfnamefont {K.}~\bibnamefont {Horodecki}},\ }\bibfield  {title} {\bibinfo {title} {Quantum entanglement},\ }\href {https://doi.org/10.1103/RevModPhys.81.865} {\bibfield  {journal} {\bibinfo  {journal} {Rev. Mod. Phys.}\ }\textbf {\bibinfo {volume} {81}},\ \bibinfo {pages} {865} (\bibinfo {year} {2009})}\BibitemShut {NoStop}%
\bibitem [{\citenamefont {G{\"u}hne}\ and\ \citenamefont {T{\'o}th}(2009)}]{guhne2009entanglement}%
  \BibitemOpen
  \bibfield  {author} {\bibinfo {author} {\bibfnamefont {O.}~\bibnamefont {G{\"u}hne}}\ and\ \bibinfo {author} {\bibfnamefont {G.}~\bibnamefont {T{\'o}th}},\ }\bibfield  {title} {\bibinfo {title} {Entanglement detection},\ }\href@noop {} {\bibfield  {journal} {\bibinfo  {journal} {Physics Reports}\ }\textbf {\bibinfo {volume} {474}},\ \bibinfo {pages} {1} (\bibinfo {year} {2009})}\BibitemShut {NoStop}%
\bibitem [{\citenamefont {Peres}(1996)}]{PhysRevLett.77.1413}%
  \BibitemOpen
  \bibfield  {author} {\bibinfo {author} {\bibfnamefont {A.}~\bibnamefont {Peres}},\ }\bibfield  {title} {\bibinfo {title} {Separability criterion for density matrices},\ }\href {https://doi.org/10.1103/PhysRevLett.77.1413} {\bibfield  {journal} {\bibinfo  {journal} {Phys. Rev. Lett.}\ }\textbf {\bibinfo {volume} {77}},\ \bibinfo {pages} {1413} (\bibinfo {year} {1996})}\BibitemShut {NoStop}%
\bibitem [{\citenamefont {Zhou}\ \emph {et~al.}(2019)\citenamefont {Zhou}, \citenamefont {Zhao}, \citenamefont {Yuan},\ and\ \citenamefont {Ma}}]{zhou2019detecting}%
  \BibitemOpen
  \bibfield  {author} {\bibinfo {author} {\bibfnamefont {Y.}~\bibnamefont {Zhou}}, \bibinfo {author} {\bibfnamefont {Q.}~\bibnamefont {Zhao}}, \bibinfo {author} {\bibfnamefont {X.}~\bibnamefont {Yuan}},\ and\ \bibinfo {author} {\bibfnamefont {X.}~\bibnamefont {Ma}},\ }\bibfield  {title} {\bibinfo {title} {Detecting multipartite entanglement structure with minimal resources},\ }\href@noop {} {\bibfield  {journal} {\bibinfo  {journal} {npj Quantum Information}\ }\textbf {\bibinfo {volume} {5}},\ \bibinfo {pages} {83} (\bibinfo {year} {2019})}\BibitemShut {NoStop}%
\bibitem [{\citenamefont {Li}\ \emph {et~al.}(2024)\citenamefont {Li}, \citenamefont {Du}, \citenamefont {Qin}, \citenamefont {Zhang}, \citenamefont {Du}, \citenamefont {Zhou},\ and\ \citenamefont {Xiao}}]{li2024entanglement}%
  \BibitemOpen
  \bibfield  {author} {\bibinfo {author} {\bibfnamefont {R.}~\bibnamefont {Li}}, \bibinfo {author} {\bibfnamefont {J.}~\bibnamefont {Du}}, \bibinfo {author} {\bibfnamefont {Z.}~\bibnamefont {Qin}}, \bibinfo {author} {\bibfnamefont {S.}~\bibnamefont {Zhang}}, \bibinfo {author} {\bibfnamefont {C.}~\bibnamefont {Du}}, \bibinfo {author} {\bibfnamefont {Y.}~\bibnamefont {Zhou}},\ and\ \bibinfo {author} {\bibfnamefont {Z.}~\bibnamefont {Xiao}},\ }\bibfield  {title} {\bibinfo {title} {Entanglement structure detection via computer vision},\ }\href@noop {} {\bibfield  {journal} {\bibinfo  {journal} {Physical Review A}\ }\textbf {\bibinfo {volume} {110}},\ \bibinfo {pages} {012448} (\bibinfo {year} {2024})}\BibitemShut {NoStop}%
\bibitem [{\citenamefont {Chen}\ \emph {et~al.}(2021)\citenamefont {Chen}, \citenamefont {Pan}, \citenamefont {Zhang},\ and\ \citenamefont {Cheng}}]{chen2021detecting}%
  \BibitemOpen
  \bibfield  {author} {\bibinfo {author} {\bibfnamefont {Y.}~\bibnamefont {Chen}}, \bibinfo {author} {\bibfnamefont {Y.}~\bibnamefont {Pan}}, \bibinfo {author} {\bibfnamefont {G.}~\bibnamefont {Zhang}},\ and\ \bibinfo {author} {\bibfnamefont {S.}~\bibnamefont {Cheng}},\ }\bibfield  {title} {\bibinfo {title} {Detecting quantum entanglement with unsupervised learning},\ }\href@noop {} {\bibfield  {journal} {\bibinfo  {journal} {Quantum Science and Technology}\ }\textbf {\bibinfo {volume} {7}},\ \bibinfo {pages} {015005} (\bibinfo {year} {2021})}\BibitemShut {NoStop}%
\bibitem [{\citenamefont {Zhang}\ \emph {et~al.}(2023)\citenamefont {Zhang}, \citenamefont {Chen},\ and\ \citenamefont {Fei}}]{zhang2023entanglement}%
  \BibitemOpen
  \bibfield  {author} {\bibinfo {author} {\bibfnamefont {L.}~\bibnamefont {Zhang}}, \bibinfo {author} {\bibfnamefont {Z.}~\bibnamefont {Chen}},\ and\ \bibinfo {author} {\bibfnamefont {S.-M.}\ \bibnamefont {Fei}},\ }\bibfield  {title} {\bibinfo {title} {Entanglement verification with deep semisupervised machine learning},\ }\href@noop {} {\bibfield  {journal} {\bibinfo  {journal} {Physical Review A}\ }\textbf {\bibinfo {volume} {108}},\ \bibinfo {pages} {022427} (\bibinfo {year} {2023})}\BibitemShut {NoStop}%
\bibitem [{\citenamefont {Paris}\ and\ \citenamefont {Rehacek}(2004)}]{paris2004quantum}%
  \BibitemOpen
  \bibfield  {author} {\bibinfo {author} {\bibfnamefont {M.}~\bibnamefont {Paris}}\ and\ \bibinfo {author} {\bibfnamefont {J.}~\bibnamefont {Rehacek}},\ }\href@noop {} {\emph {\bibinfo {title} {Quantum state estimation}}},\ Vol.\ \bibinfo {volume} {649}\ (\bibinfo  {publisher} {Springer Science \& Business Media},\ \bibinfo {year} {2004})\BibitemShut {NoStop}%
\bibitem [{\citenamefont {Ahmed}\ \emph {et~al.}(2021)\citenamefont {Ahmed}, \citenamefont {S{\'a}nchez~Mu{\~n}oz}, \citenamefont {Nori},\ and\ \citenamefont {Kockum}}]{ahmed2021quantum}%
  \BibitemOpen
  \bibfield  {author} {\bibinfo {author} {\bibfnamefont {S.}~\bibnamefont {Ahmed}}, \bibinfo {author} {\bibfnamefont {C.}~\bibnamefont {S{\'a}nchez~Mu{\~n}oz}}, \bibinfo {author} {\bibfnamefont {F.}~\bibnamefont {Nori}},\ and\ \bibinfo {author} {\bibfnamefont {A.~F.}\ \bibnamefont {Kockum}},\ }\bibfield  {title} {\bibinfo {title} {Quantum state tomography with conditional generative adversarial networks},\ }\href@noop {} {\bibfield  {journal} {\bibinfo  {journal} {Physical review letters}\ }\textbf {\bibinfo {volume} {127}},\ \bibinfo {pages} {140502} (\bibinfo {year} {2021})}\BibitemShut {NoStop}%
\bibitem [{\citenamefont {Xin}\ \emph {et~al.}(2017)\citenamefont {Xin}, \citenamefont {Lu}, \citenamefont {Klassen}, \citenamefont {Yu}, \citenamefont {Ji}, \citenamefont {Chen}, \citenamefont {Ma}, \citenamefont {Long}, \citenamefont {Zeng},\ and\ \citenamefont {Laflamme}}]{xin2017quantum}%
  \BibitemOpen
  \bibfield  {author} {\bibinfo {author} {\bibfnamefont {T.}~\bibnamefont {Xin}}, \bibinfo {author} {\bibfnamefont {D.}~\bibnamefont {Lu}}, \bibinfo {author} {\bibfnamefont {J.}~\bibnamefont {Klassen}}, \bibinfo {author} {\bibfnamefont {N.}~\bibnamefont {Yu}}, \bibinfo {author} {\bibfnamefont {Z.}~\bibnamefont {Ji}}, \bibinfo {author} {\bibfnamefont {J.}~\bibnamefont {Chen}}, \bibinfo {author} {\bibfnamefont {X.}~\bibnamefont {Ma}}, \bibinfo {author} {\bibfnamefont {G.}~\bibnamefont {Long}}, \bibinfo {author} {\bibfnamefont {B.}~\bibnamefont {Zeng}},\ and\ \bibinfo {author} {\bibfnamefont {R.}~\bibnamefont {Laflamme}},\ }\bibfield  {title} {\bibinfo {title} {Quantum state tomography via reduced density matrices},\ }\href@noop {} {\bibfield  {journal} {\bibinfo  {journal} {Physical review letters}\ }\textbf {\bibinfo {volume} {118}},\ \bibinfo {pages} {020401} (\bibinfo {year} {2017})}\BibitemShut {NoStop}%
\bibitem [{\citenamefont {Gross}\ \emph {et~al.}(2010)\citenamefont {Gross}, \citenamefont {Liu}, \citenamefont {Flammia}, \citenamefont {Becker},\ and\ \citenamefont {Eisert}}]{gross2010quantum}%
  \BibitemOpen
  \bibfield  {author} {\bibinfo {author} {\bibfnamefont {D.}~\bibnamefont {Gross}}, \bibinfo {author} {\bibfnamefont {Y.-K.}\ \bibnamefont {Liu}}, \bibinfo {author} {\bibfnamefont {S.~T.}\ \bibnamefont {Flammia}}, \bibinfo {author} {\bibfnamefont {S.}~\bibnamefont {Becker}},\ and\ \bibinfo {author} {\bibfnamefont {J.}~\bibnamefont {Eisert}},\ }\bibfield  {title} {\bibinfo {title} {Quantum state tomography via compressed sensing},\ }\href@noop {} {\bibfield  {journal} {\bibinfo  {journal} {Physical review letters}\ }\textbf {\bibinfo {volume} {105}},\ \bibinfo {pages} {150401} (\bibinfo {year} {2010})}\BibitemShut {NoStop}%
\bibitem [{\citenamefont {Altepeter}\ \emph {et~al.}(2004)\citenamefont {Altepeter}, \citenamefont {James},\ and\ \citenamefont {Kwiat}}]{Altepeter2004}%
  \BibitemOpen
  \bibfield  {author} {\bibinfo {author} {\bibfnamefont {J.~B.}\ \bibnamefont {Altepeter}}, \bibinfo {author} {\bibfnamefont {D.~F.}\ \bibnamefont {James}},\ and\ \bibinfo {author} {\bibfnamefont {P.~G.}\ \bibnamefont {Kwiat}},\ }\bibinfo {title} {4 qubit quantum state tomography},\ in\ \href {https://doi.org/10.1007/978-3-540-44481-7_4} {\emph {\bibinfo {booktitle} {Quantum State Estimation}}},\ \bibinfo {editor} {edited by\ \bibinfo {editor} {\bibfnamefont {M.}~\bibnamefont {Paris}}\ and\ \bibinfo {editor} {\bibfnamefont {J.}~\bibnamefont {{\v{R}}eh{\'a}{\v{c}}ek}}}\ (\bibinfo  {publisher} {Springer Berlin Heidelberg},\ \bibinfo {address} {Berlin, Heidelberg},\ \bibinfo {year} {2004})\ pp.\ \bibinfo {pages} {113--145}\BibitemShut {NoStop}%
\bibitem [{\citenamefont {Torlai}\ \emph {et~al.}(2018)\citenamefont {Torlai}, \citenamefont {Mazzola}, \citenamefont {Carrasquilla}, \citenamefont {Troyer}, \citenamefont {Melko},\ and\ \citenamefont {Carleo}}]{torlai2018neural}%
  \BibitemOpen
  \bibfield  {author} {\bibinfo {author} {\bibfnamefont {G.}~\bibnamefont {Torlai}}, \bibinfo {author} {\bibfnamefont {G.}~\bibnamefont {Mazzola}}, \bibinfo {author} {\bibfnamefont {J.}~\bibnamefont {Carrasquilla}}, \bibinfo {author} {\bibfnamefont {M.}~\bibnamefont {Troyer}}, \bibinfo {author} {\bibfnamefont {R.}~\bibnamefont {Melko}},\ and\ \bibinfo {author} {\bibfnamefont {G.}~\bibnamefont {Carleo}},\ }\bibfield  {title} {\bibinfo {title} {Neural-network quantum state tomography},\ }\href@noop {} {\bibfield  {journal} {\bibinfo  {journal} {Nature physics}\ }\textbf {\bibinfo {volume} {14}},\ \bibinfo {pages} {447} (\bibinfo {year} {2018})}\BibitemShut {NoStop}%
\bibitem [{\citenamefont {Gao}\ \emph {et~al.}(2018)\citenamefont {Gao}, \citenamefont {Qiao}, \citenamefont {Jiao}, \citenamefont {Ma}, \citenamefont {Hu}, \citenamefont {Ren}, \citenamefont {Yang}, \citenamefont {Tang}, \citenamefont {Yung},\ and\ \citenamefont {Jin}}]{gao2018experimental}%
  \BibitemOpen
  \bibfield  {author} {\bibinfo {author} {\bibfnamefont {J.}~\bibnamefont {Gao}}, \bibinfo {author} {\bibfnamefont {L.-F.}\ \bibnamefont {Qiao}}, \bibinfo {author} {\bibfnamefont {Z.-Q.}\ \bibnamefont {Jiao}}, \bibinfo {author} {\bibfnamefont {Y.-C.}\ \bibnamefont {Ma}}, \bibinfo {author} {\bibfnamefont {C.-Q.}\ \bibnamefont {Hu}}, \bibinfo {author} {\bibfnamefont {R.-J.}\ \bibnamefont {Ren}}, \bibinfo {author} {\bibfnamefont {A.-L.}\ \bibnamefont {Yang}}, \bibinfo {author} {\bibfnamefont {H.}~\bibnamefont {Tang}}, \bibinfo {author} {\bibfnamefont {M.-H.}\ \bibnamefont {Yung}},\ and\ \bibinfo {author} {\bibfnamefont {X.-M.}\ \bibnamefont {Jin}},\ }\bibfield  {title} {\bibinfo {title} {Experimental machine learning of quantum states},\ }\href@noop {} {\bibfield  {journal} {\bibinfo  {journal} {Physical review letters}\ }\textbf {\bibinfo {volume} {120}},\ \bibinfo {pages} {240501} (\bibinfo {year} {2018})}\BibitemShut {NoStop}%
\bibitem [{\citenamefont {Carleo}\ and\ \citenamefont {Troyer}(2017)}]{carleo2017solving}%
  \BibitemOpen
  \bibfield  {author} {\bibinfo {author} {\bibfnamefont {G.}~\bibnamefont {Carleo}}\ and\ \bibinfo {author} {\bibfnamefont {M.}~\bibnamefont {Troyer}},\ }\bibfield  {title} {\bibinfo {title} {Solving the quantum many-body problem with artificial neural networks},\ }\href@noop {} {\bibfield  {journal} {\bibinfo  {journal} {Science}\ }\textbf {\bibinfo {volume} {355}},\ \bibinfo {pages} {602} (\bibinfo {year} {2017})}\BibitemShut {NoStop}%
\bibitem [{\citenamefont {Carrasquilla}(2020)}]{carrasquilla2020machine}%
  \BibitemOpen
  \bibfield  {author} {\bibinfo {author} {\bibfnamefont {J.}~\bibnamefont {Carrasquilla}},\ }\bibfield  {title} {\bibinfo {title} {Machine learning for quantum matter},\ }\href@noop {} {\bibfield  {journal} {\bibinfo  {journal} {Advances in Physics: X}\ }\textbf {\bibinfo {volume} {5}},\ \bibinfo {pages} {1797528} (\bibinfo {year} {2020})}\BibitemShut {NoStop}%
\bibitem [{\citenamefont {Dunjko}\ and\ \citenamefont {Briegel}(2018)}]{dunjko2018machine}%
  \BibitemOpen
  \bibfield  {author} {\bibinfo {author} {\bibfnamefont {V.}~\bibnamefont {Dunjko}}\ and\ \bibinfo {author} {\bibfnamefont {H.~J.}\ \bibnamefont {Briegel}},\ }\bibfield  {title} {\bibinfo {title} {Machine learning \& artificial intelligence in the quantum domain: a review of recent progress},\ }\href@noop {} {\bibfield  {journal} {\bibinfo  {journal} {Reports on Progress in Physics}\ }\textbf {\bibinfo {volume} {81}},\ \bibinfo {pages} {074001} (\bibinfo {year} {2018})}\BibitemShut {NoStop}%
\bibitem [{\citenamefont {Li}\ \emph {et~al.}(2018)\citenamefont {Li}, \citenamefont {Yang},\ and\ \citenamefont {Zhang}}]{li2018survey}%
  \BibitemOpen
  \bibfield  {author} {\bibinfo {author} {\bibfnamefont {Y.}~\bibnamefont {Li}}, \bibinfo {author} {\bibfnamefont {M.}~\bibnamefont {Yang}},\ and\ \bibinfo {author} {\bibfnamefont {Z.}~\bibnamefont {Zhang}},\ }\bibfield  {title} {\bibinfo {title} {A survey of multi-view representation learning},\ }\href@noop {} {\bibfield  {journal} {\bibinfo  {journal} {IEEE transactions on knowledge and data engineering}\ }\textbf {\bibinfo {volume} {31}},\ \bibinfo {pages} {1863} (\bibinfo {year} {2018})}\BibitemShut {NoStop}%
\bibitem [{\citenamefont {Wang}\ \emph {et~al.}(2015)\citenamefont {Wang}, \citenamefont {Arora}, \citenamefont {Livescu},\ and\ \citenamefont {Bilmes}}]{wang2015deep}%
  \BibitemOpen
  \bibfield  {author} {\bibinfo {author} {\bibfnamefont {W.}~\bibnamefont {Wang}}, \bibinfo {author} {\bibfnamefont {R.}~\bibnamefont {Arora}}, \bibinfo {author} {\bibfnamefont {K.}~\bibnamefont {Livescu}},\ and\ \bibinfo {author} {\bibfnamefont {J.}~\bibnamefont {Bilmes}},\ }\bibfield  {title} {\bibinfo {title} {On deep multi-view representation learning},\ }in\ \href@noop {} {\emph {\bibinfo {booktitle} {International conference on machine learning}}}\ (\bibinfo {organization} {PMLR},\ \bibinfo {year} {2015})\ pp.\ \bibinfo {pages} {1083--1092}\BibitemShut {NoStop}%
\bibitem [{\citenamefont {Lewenstein}\ \emph {et~al.}(2000)\citenamefont {Lewenstein}, \citenamefont {Kraus}, \citenamefont {Cirac},\ and\ \citenamefont {Horodecki}}]{lewenstein2000optimization}%
  \BibitemOpen
  \bibfield  {author} {\bibinfo {author} {\bibfnamefont {M.}~\bibnamefont {Lewenstein}}, \bibinfo {author} {\bibfnamefont {B.}~\bibnamefont {Kraus}}, \bibinfo {author} {\bibfnamefont {J.~I.}\ \bibnamefont {Cirac}},\ and\ \bibinfo {author} {\bibfnamefont {P.}~\bibnamefont {Horodecki}},\ }\bibfield  {title} {\bibinfo {title} {Optimization of entanglement witnesses},\ }\href@noop {} {\bibfield  {journal} {\bibinfo  {journal} {Physical Review A}\ }\textbf {\bibinfo {volume} {62}},\ \bibinfo {pages} {052310} (\bibinfo {year} {2000})}\BibitemShut {NoStop}%
\bibitem [{\citenamefont {T{\'o}th}\ and\ \citenamefont {G{\"u}hne}(2005)}]{toth2005detecting}%
  \BibitemOpen
  \bibfield  {author} {\bibinfo {author} {\bibfnamefont {G.}~\bibnamefont {T{\'o}th}}\ and\ \bibinfo {author} {\bibfnamefont {O.}~\bibnamefont {G{\"u}hne}},\ }\bibfield  {title} {\bibinfo {title} {Detecting genuine multipartite entanglement with two local measurements},\ }\href@noop {} {\bibfield  {journal} {\bibinfo  {journal} {Physical review letters}\ }\textbf {\bibinfo {volume} {94}},\ \bibinfo {pages} {060501} (\bibinfo {year} {2005})}\BibitemShut {NoStop}%
\bibitem [{\citenamefont {Ou}\ and\ \citenamefont {Murphey}(2007)}]{OU20074}%
  \BibitemOpen
  \bibfield  {author} {\bibinfo {author} {\bibfnamefont {G.}~\bibnamefont {Ou}}\ and\ \bibinfo {author} {\bibfnamefont {Y.~L.}\ \bibnamefont {Murphey}},\ }\bibfield  {title} {\bibinfo {title} {Multi-class pattern classification using neural networks},\ }\href {https://doi.org/https://doi.org/10.1016/j.patcog.2006.04.041} {\bibfield  {journal} {\bibinfo  {journal} {Pattern Recognition}\ }\textbf {\bibinfo {volume} {40}},\ \bibinfo {pages} {4} (\bibinfo {year} {2007})}\BibitemShut {NoStop}%
\bibitem [{\citenamefont {Bergholm}\ \emph {et~al.}(2018)\citenamefont {Bergholm}, \citenamefont {Izaac}, \citenamefont {Schuld}, \citenamefont {Gogolin}, \citenamefont {Ahmed}, \citenamefont {Ajith}, \citenamefont {Alam}, \citenamefont {Alonso-Linaje}, \citenamefont {AkashNarayanan}, \citenamefont {Asadi} \emph {et~al.}}]{bergholm2018pennylane}%
  \BibitemOpen
  \bibfield  {author} {\bibinfo {author} {\bibfnamefont {V.}~\bibnamefont {Bergholm}}, \bibinfo {author} {\bibfnamefont {J.}~\bibnamefont {Izaac}}, \bibinfo {author} {\bibfnamefont {M.}~\bibnamefont {Schuld}}, \bibinfo {author} {\bibfnamefont {C.}~\bibnamefont {Gogolin}}, \bibinfo {author} {\bibfnamefont {S.}~\bibnamefont {Ahmed}}, \bibinfo {author} {\bibfnamefont {V.}~\bibnamefont {Ajith}}, \bibinfo {author} {\bibfnamefont {M.~S.}\ \bibnamefont {Alam}}, \bibinfo {author} {\bibfnamefont {G.}~\bibnamefont {Alonso-Linaje}}, \bibinfo {author} {\bibfnamefont {B.}~\bibnamefont {AkashNarayanan}}, \bibinfo {author} {\bibfnamefont {A.}~\bibnamefont {Asadi}}, \emph {et~al.},\ }\bibfield  {title} {\bibinfo {title} {Pennylane: Automatic differentiation of hybrid quantum-classical computations},\ }\href@noop {} {\bibfield  {journal} {\bibinfo  {journal} {arXiv preprint arXiv:1811.04968}\ } (\bibinfo {year} {2018})}\BibitemShut {NoStop}%
\bibitem [{Pen()}]{PennylaneWEB}%
  \BibitemOpen
  \href {https://pennylane.ai/} {\bibinfo {title} {Pennylane tutorial}},\ \bibinfo {note} {accessed: [Insert Access Date]}\BibitemShut {NoStop}%
\bibitem [{\citenamefont {Hahnloser}\ \emph {et~al.}(2000)\citenamefont {Hahnloser}, \citenamefont {Sarpeshkar}, \citenamefont {Mahowald}, \citenamefont {Douglas},\ and\ \citenamefont {Seung}}]{hahnloser2000digital}%
  \BibitemOpen
  \bibfield  {author} {\bibinfo {author} {\bibfnamefont {R.~H.}\ \bibnamefont {Hahnloser}}, \bibinfo {author} {\bibfnamefont {R.}~\bibnamefont {Sarpeshkar}}, \bibinfo {author} {\bibfnamefont {M.~A.}\ \bibnamefont {Mahowald}}, \bibinfo {author} {\bibfnamefont {R.~J.}\ \bibnamefont {Douglas}},\ and\ \bibinfo {author} {\bibfnamefont {H.~S.}\ \bibnamefont {Seung}},\ }\bibfield  {title} {\bibinfo {title} {Digital selection and analogue amplification coexist in a cortex-inspired silicon circuit},\ }\href@noop {} {\bibfield  {journal} {\bibinfo  {journal} {nature}\ }\textbf {\bibinfo {volume} {405}},\ \bibinfo {pages} {947} (\bibinfo {year} {2000})}\BibitemShut {NoStop}%
\bibitem [{\citenamefont {Zhang}\ and\ \citenamefont {Sabuncu}(2018)}]{zhang2018generalized}%
  \BibitemOpen
  \bibfield  {author} {\bibinfo {author} {\bibfnamefont {Z.}~\bibnamefont {Zhang}}\ and\ \bibinfo {author} {\bibfnamefont {M.}~\bibnamefont {Sabuncu}},\ }\bibfield  {title} {\bibinfo {title} {Generalized cross entropy loss for training deep neural networks with noisy labels},\ }\href@noop {} {\bibfield  {journal} {\bibinfo  {journal} {Advances in neural information processing systems}\ }\textbf {\bibinfo {volume} {31}} (\bibinfo {year} {2018})}\BibitemShut {NoStop}%
\bibitem [{\citenamefont {Khosla}\ \emph {et~al.}(2020)\citenamefont {Khosla}, \citenamefont {Teterwak}, \citenamefont {Wang}, \citenamefont {Sarna}, \citenamefont {Tian}, \citenamefont {Isola}, \citenamefont {Maschinot}, \citenamefont {Liu},\ and\ \citenamefont {Krishnan}}]{khosla2020supervised}%
  \BibitemOpen
  \bibfield  {author} {\bibinfo {author} {\bibfnamefont {P.}~\bibnamefont {Khosla}}, \bibinfo {author} {\bibfnamefont {P.}~\bibnamefont {Teterwak}}, \bibinfo {author} {\bibfnamefont {C.}~\bibnamefont {Wang}}, \bibinfo {author} {\bibfnamefont {A.}~\bibnamefont {Sarna}}, \bibinfo {author} {\bibfnamefont {Y.}~\bibnamefont {Tian}}, \bibinfo {author} {\bibfnamefont {P.}~\bibnamefont {Isola}}, \bibinfo {author} {\bibfnamefont {A.}~\bibnamefont {Maschinot}}, \bibinfo {author} {\bibfnamefont {C.}~\bibnamefont {Liu}},\ and\ \bibinfo {author} {\bibfnamefont {D.}~\bibnamefont {Krishnan}},\ }\bibfield  {title} {\bibinfo {title} {Supervised contrastive learning},\ }\href@noop {} {\bibfield  {journal} {\bibinfo  {journal} {Advances in neural information processing systems}\ }\textbf {\bibinfo {volume} {33}},\ \bibinfo {pages} {18661} (\bibinfo {year} {2020})}\BibitemShut {NoStop}%
\bibitem [{\citenamefont {Nguyen}\ and\ \citenamefont {Bai}(2010)}]{nguyen2010cosine}%
  \BibitemOpen
  \bibfield  {author} {\bibinfo {author} {\bibfnamefont {H.~V.}\ \bibnamefont {Nguyen}}\ and\ \bibinfo {author} {\bibfnamefont {L.}~\bibnamefont {Bai}},\ }\bibfield  {title} {\bibinfo {title} {Cosine similarity metric learning for face verification},\ }in\ \href@noop {} {\emph {\bibinfo {booktitle} {Asian conference on computer vision}}}\ (\bibinfo {organization} {Springer},\ \bibinfo {year} {2010})\ pp.\ \bibinfo {pages} {709--720}\BibitemShut {NoStop}%
\bibitem [{\citenamefont {Kingma}(2014)}]{kingma2014adam}%
  \BibitemOpen
  \bibfield  {author} {\bibinfo {author} {\bibfnamefont {D.~P.}\ \bibnamefont {Kingma}},\ }\bibfield  {title} {\bibinfo {title} {Adam: A method for stochastic optimization},\ }\href@noop {} {\bibfield  {journal} {\bibinfo  {journal} {arXiv preprint arXiv:1412.6980}\ } (\bibinfo {year} {2014})}\BibitemShut {NoStop}%
\bibitem [{\citenamefont {Rodionov}\ \emph {et~al.}(2014)\citenamefont {Rodionov}, \citenamefont {Veitia}, \citenamefont {Barends}, \citenamefont {Kelly}, \citenamefont {Sank}, \citenamefont {Wenner}, \citenamefont {Martinis}, \citenamefont {Kosut},\ and\ \citenamefont {Korotkov}}]{rodionov2014compressed}%
  \BibitemOpen
  \bibfield  {author} {\bibinfo {author} {\bibfnamefont {A.~V.}\ \bibnamefont {Rodionov}}, \bibinfo {author} {\bibfnamefont {A.}~\bibnamefont {Veitia}}, \bibinfo {author} {\bibfnamefont {R.}~\bibnamefont {Barends}}, \bibinfo {author} {\bibfnamefont {J.}~\bibnamefont {Kelly}}, \bibinfo {author} {\bibfnamefont {D.}~\bibnamefont {Sank}}, \bibinfo {author} {\bibfnamefont {J.}~\bibnamefont {Wenner}}, \bibinfo {author} {\bibfnamefont {J.~M.}\ \bibnamefont {Martinis}}, \bibinfo {author} {\bibfnamefont {R.~L.}\ \bibnamefont {Kosut}},\ and\ \bibinfo {author} {\bibfnamefont {A.~N.}\ \bibnamefont {Korotkov}},\ }\bibfield  {title} {\bibinfo {title} {Compressed sensing quantum process tomography for superconducting quantum gates},\ }\href@noop {} {\bibfield  {journal} {\bibinfo  {journal} {Physical Review B}\ }\textbf {\bibinfo {volume} {90}},\ \bibinfo {pages} {144504} (\bibinfo {year} {2014})}\BibitemShut {NoStop}%
\bibitem [{\citenamefont {Gebhart}\ \emph {et~al.}(2023)\citenamefont {Gebhart}, \citenamefont {Santagati}, \citenamefont {Gentile}, \citenamefont {Gauger}, \citenamefont {Craig}, \citenamefont {Ares}, \citenamefont {Banchi}, \citenamefont {Marquardt}, \citenamefont {Pezz{\`e}},\ and\ \citenamefont {Bonato}}]{gebhart2023learning}%
  \BibitemOpen
  \bibfield  {author} {\bibinfo {author} {\bibfnamefont {V.}~\bibnamefont {Gebhart}}, \bibinfo {author} {\bibfnamefont {R.}~\bibnamefont {Santagati}}, \bibinfo {author} {\bibfnamefont {A.~A.}\ \bibnamefont {Gentile}}, \bibinfo {author} {\bibfnamefont {E.~M.}\ \bibnamefont {Gauger}}, \bibinfo {author} {\bibfnamefont {D.}~\bibnamefont {Craig}}, \bibinfo {author} {\bibfnamefont {N.}~\bibnamefont {Ares}}, \bibinfo {author} {\bibfnamefont {L.}~\bibnamefont {Banchi}}, \bibinfo {author} {\bibfnamefont {F.}~\bibnamefont {Marquardt}}, \bibinfo {author} {\bibfnamefont {L.}~\bibnamefont {Pezz{\`e}}},\ and\ \bibinfo {author} {\bibfnamefont {C.}~\bibnamefont {Bonato}},\ }\bibfield  {title} {\bibinfo {title} {Learning quantum systems},\ }\href@noop {} {\bibfield  {journal} {\bibinfo  {journal} {Nature Reviews Physics}\ }\textbf {\bibinfo {volume} {5}},\ \bibinfo {pages} {141} (\bibinfo {year} {2023})}\BibitemShut {NoStop}%
\bibitem [{\citenamefont {Smith}\ \emph {et~al.}(2021)\citenamefont {Smith}, \citenamefont {Gray},\ and\ \citenamefont {Kim}}]{PRXQuantum.2.020348}%
  \BibitemOpen
  \bibfield  {author} {\bibinfo {author} {\bibfnamefont {A.~W.~R.}\ \bibnamefont {Smith}}, \bibinfo {author} {\bibfnamefont {J.}~\bibnamefont {Gray}},\ and\ \bibinfo {author} {\bibfnamefont {M.~S.}\ \bibnamefont {Kim}},\ }\bibfield  {title} {\bibinfo {title} {Efficient quantum state sample tomography with basis-dependent neural networks},\ }\href {https://doi.org/10.1103/PRXQuantum.2.020348} {\bibfield  {journal} {\bibinfo  {journal} {PRX Quantum}\ }\textbf {\bibinfo {volume} {2}},\ \bibinfo {pages} {020348} (\bibinfo {year} {2021})}\BibitemShut {NoStop}%
\bibitem [{\citenamefont {Fei}(2017)}]{fei2017adaptive}%
  \BibitemOpen
  \bibfield  {author} {\bibinfo {author} {\bibfnamefont {S.-M.}\ \bibnamefont {Fei}},\ }\bibfield  {title} {\bibinfo {title} {Adaptive-measurement based quantum tomography},\ }\href@noop {} {\bibfield  {journal} {\bibinfo  {journal} {Science China. Physics, Mechanics \& Astronomy}\ }\textbf {\bibinfo {volume} {60}},\ \bibinfo {pages} {020331} (\bibinfo {year} {2017})}\BibitemShut {NoStop}%
\bibitem [{\citenamefont {Quek}\ \emph {et~al.}(2021)\citenamefont {Quek}, \citenamefont {Fort},\ and\ \citenamefont {Ng}}]{quek2021adaptive}%
  \BibitemOpen
  \bibfield  {author} {\bibinfo {author} {\bibfnamefont {Y.}~\bibnamefont {Quek}}, \bibinfo {author} {\bibfnamefont {S.}~\bibnamefont {Fort}},\ and\ \bibinfo {author} {\bibfnamefont {H.~K.}\ \bibnamefont {Ng}},\ }\bibfield  {title} {\bibinfo {title} {Adaptive quantum state tomography with neural networks},\ }\href@noop {} {\bibfield  {journal} {\bibinfo  {journal} {npj Quantum Information}\ }\textbf {\bibinfo {volume} {7}},\ \bibinfo {pages} {105} (\bibinfo {year} {2021})}\BibitemShut {NoStop}%
\bibitem [{\citenamefont {Hwang}\ \emph {et~al.}(2023)\citenamefont {Hwang}, \citenamefont {Choi},\ and\ \citenamefont {Kim}}]{PhysRevApplied.20.064007}%
  \BibitemOpen
  \bibfield  {author} {\bibinfo {author} {\bibfnamefont {H.}~\bibnamefont {Hwang}}, \bibinfo {author} {\bibfnamefont {J.}~\bibnamefont {Choi}},\ and\ \bibinfo {author} {\bibfnamefont {E.}~\bibnamefont {Kim}},\ }\bibfield  {title} {\bibinfo {title} {Adaptive quantum tomography in an indistinct measurement system with superconducting circuits},\ }\href {https://doi.org/10.1103/PhysRevApplied.20.064007} {\bibfield  {journal} {\bibinfo  {journal} {Phys. Rev. Appl.}\ }\textbf {\bibinfo {volume} {20}},\ \bibinfo {pages} {064007} (\bibinfo {year} {2023})}\BibitemShut {NoStop}%
\bibitem [{\citenamefont {Martin}\ \emph {et~al.}(2017)\citenamefont {Martin}, \citenamefont {Guerreiro}, \citenamefont {Tiranov}, \citenamefont {Designolle}, \citenamefont {Fr{\"o}wis}, \citenamefont {Brunner}, \citenamefont {Huber},\ and\ \citenamefont {Gisin}}]{martin2017quantifying}%
  \BibitemOpen
  \bibfield  {author} {\bibinfo {author} {\bibfnamefont {A.}~\bibnamefont {Martin}}, \bibinfo {author} {\bibfnamefont {T.}~\bibnamefont {Guerreiro}}, \bibinfo {author} {\bibfnamefont {A.}~\bibnamefont {Tiranov}}, \bibinfo {author} {\bibfnamefont {S.}~\bibnamefont {Designolle}}, \bibinfo {author} {\bibfnamefont {F.}~\bibnamefont {Fr{\"o}wis}}, \bibinfo {author} {\bibfnamefont {N.}~\bibnamefont {Brunner}}, \bibinfo {author} {\bibfnamefont {M.}~\bibnamefont {Huber}},\ and\ \bibinfo {author} {\bibfnamefont {N.}~\bibnamefont {Gisin}},\ }\bibfield  {title} {\bibinfo {title} {Quantifying photonic high-dimensional entanglement},\ }\href@noop {} {\bibfield  {journal} {\bibinfo  {journal} {Physical review letters}\ }\textbf {\bibinfo {volume} {118}},\ \bibinfo {pages} {110501} (\bibinfo {year} {2017})}\BibitemShut {NoStop}%
\bibitem [{\citenamefont {Chang}\ \emph {et~al.}(2021)\citenamefont {Chang}, \citenamefont {Cheng}, \citenamefont {Sarihan}, \citenamefont {Vinod}, \citenamefont {Lee}, \citenamefont {Zhong}, \citenamefont {Gong}, \citenamefont {Xie}, \citenamefont {Shapiro}, \citenamefont {Wong} \emph {et~al.}}]{chang2021648}%
  \BibitemOpen
  \bibfield  {author} {\bibinfo {author} {\bibfnamefont {K.-C.}\ \bibnamefont {Chang}}, \bibinfo {author} {\bibfnamefont {X.}~\bibnamefont {Cheng}}, \bibinfo {author} {\bibfnamefont {M.~C.}\ \bibnamefont {Sarihan}}, \bibinfo {author} {\bibfnamefont {A.~K.}\ \bibnamefont {Vinod}}, \bibinfo {author} {\bibfnamefont {Y.~S.}\ \bibnamefont {Lee}}, \bibinfo {author} {\bibfnamefont {T.}~\bibnamefont {Zhong}}, \bibinfo {author} {\bibfnamefont {Y.-X.}\ \bibnamefont {Gong}}, \bibinfo {author} {\bibfnamefont {Z.}~\bibnamefont {Xie}}, \bibinfo {author} {\bibfnamefont {J.~H.}\ \bibnamefont {Shapiro}}, \bibinfo {author} {\bibfnamefont {F.~N.}\ \bibnamefont {Wong}}, \emph {et~al.},\ }\bibfield  {title} {\bibinfo {title} {648 hilbert-space dimensionality in a biphoton frequency comb: entanglement of formation and schmidt mode decomposition},\ }\href@noop {} {\bibfield  {journal} {\bibinfo  {journal} {npj Quantum Information}\ }\textbf {\bibinfo {volume} {7}},\ \bibinfo {pages} {48} (\bibinfo {year} {2021})}\BibitemShut
  {NoStop}%
\bibitem [{\citenamefont {Friis}\ \emph {et~al.}(2019)\citenamefont {Friis}, \citenamefont {Vitagliano}, \citenamefont {Malik},\ and\ \citenamefont {Huber}}]{friis2019entanglement}%
  \BibitemOpen
  \bibfield  {author} {\bibinfo {author} {\bibfnamefont {N.}~\bibnamefont {Friis}}, \bibinfo {author} {\bibfnamefont {G.}~\bibnamefont {Vitagliano}}, \bibinfo {author} {\bibfnamefont {M.}~\bibnamefont {Malik}},\ and\ \bibinfo {author} {\bibfnamefont {M.}~\bibnamefont {Huber}},\ }\bibfield  {title} {\bibinfo {title} {Entanglement certification from theory to experiment},\ }\href@noop {} {\bibfield  {journal} {\bibinfo  {journal} {Nature Reviews Physics}\ }\textbf {\bibinfo {volume} {1}},\ \bibinfo {pages} {72} (\bibinfo {year} {2019})}\BibitemShut {NoStop}%
\bibitem [{\citenamefont {Erhard}\ \emph {et~al.}(2020)\citenamefont {Erhard}, \citenamefont {Krenn},\ and\ \citenamefont {Zeilinger}}]{erhard2020advances}%
  \BibitemOpen
  \bibfield  {author} {\bibinfo {author} {\bibfnamefont {M.}~\bibnamefont {Erhard}}, \bibinfo {author} {\bibfnamefont {M.}~\bibnamefont {Krenn}},\ and\ \bibinfo {author} {\bibfnamefont {A.}~\bibnamefont {Zeilinger}},\ }\bibfield  {title} {\bibinfo {title} {Advances in high-dimensional quantum entanglement},\ }\href@noop {} {\bibfield  {journal} {\bibinfo  {journal} {Nature Reviews Physics}\ }\textbf {\bibinfo {volume} {2}},\ \bibinfo {pages} {365} (\bibinfo {year} {2020})}\BibitemShut {NoStop}%
\end{thebibliography}%

\end{document}